\documentclass[journal,twocolumns]{IEEEtran}

\usepackage{amsfonts}
\usepackage{times}
\usepackage{graphicx}
\usepackage{epstopdf}
\usepackage{latexsym}
\usepackage{dsfont}
\usepackage{amssymb}
\usepackage{amsmath}
\usepackage{cite}
\usepackage{verbatim}

\newcommand{\figref}[1]{{Fig.}~\ref{#1}}

\newcommand{\fig}[1]{Fig.\ \ref{#1}}

\def\bb0{{\mathbb{0}}}

\def\bb{{\mathbf{b}}}
\def\bc{{\mathbf{c}}}
\def\bd{{\mathbf{d}}}
\def\bee{{\mathbf{e}}}

\def\bh{{\mathbf{h}}}

\def\bn{{\mathbf{n}}}

\def\bp{{\mathbf{p}}}

\def\br{{\mathbf{r}}}

\def\bt{{\mathbf{t}}}

\def\bv{{\mathbf{v}}}

\def\b0{{\mathbf{0}}}

\def\bE{{\mathbf{E}}}

\def\bH{{\mathbf{H}}}

\def\bS{{\mathbf{S}}}
\def\bT{{\mathbf{T}}}

\def\bbC{{\mathbb{C}}}

\def\bbE{{\mathbb{E}}}

\def\bbR{{\mathbb{R}}}

\def\cB{\mathcal{B}}

\def\cD{\mathcal{D}}
\def\cE{\mathcal{E}}
\def\cF{\mathcal{F}}
\def\cG{\mathcal{G}}
\def\cH{\mathcal{H}}
\def\cI{\mathcal{I}}

\def\cM{\mathcal{M}}
\def\cN{\mathcal{N}}
\def\cO{\mathcal{O}}

\def\cV{\mathcal{V}}

\def\sf0{{\mathsf{0}}}

\usepackage{graphicx}
\usepackage{epstopdf}
\usepackage{etoolbox} 

\usepackage{enumerate}
\usepackage{algorithm}
\usepackage{amsmath}
\usepackage{mathrsfs}
\usepackage{float}
\usepackage{color}
\usepackage{makeidx}
\usepackage{bm}
\usepackage{cleveref}
\usepackage{url}
\usepackage{steinmetz}
\usepackage{varwidth}
\usepackage{soul}
\usepackage{bbm}
\usepackage{empheq}
\usepackage{mathtools}
\usepackage{cite}
\usepackage{balance}
\usepackage{subfigure}

\newcommand{\sref}[1]{{Section}~\ref{#1}}

\def \rm {\mathrm}

\usepackage{amsfonts}
\usepackage{booktabs}
\usepackage{siunitx}

\begin{document}
	\title{Learnable Wireless Digital Twins: Reconstructing Electromagnetic Field with Neural Representations}
	\author{ Shuaifeng Jiang\textsuperscript{1},  Qi Qu\textsuperscript{2}, Xiaqing Pan\textsuperscript{2}, Abhishek Agrawal\textsuperscript{2}, Richard Newcombe\textsuperscript{2}, and Ahmed Alkhateeb\textsuperscript{1} \\
		\vspace{10pt}
		\textsuperscript{1} Wireless Intelligence Lab, Arizona State University, USA\\
		\textsuperscript{2} Meta Reality Lab, USA
		\thanks{{Shuaifeng Jiang and Ahmed Alkhateeb are with the School of Electrical, Computer, and Energy Engineering at Arizona State University - Email: \{s.jiang, alkhateeb\}@asu.edu. Qi Qu, Xiaqing Pan, Abhishek Agrawal, and Richard Newcombe are with Meta Reality Lab - Email: \{qqu, xiaqingp, abhishekag, newcombe\}@meta.com.}}} 
	\maketitle

\begin{abstract}
	Fully harvesting the gain of multiple-input and multiple-output (MIMO) requires accurate channel information. However, conventional channel acquisition methods mainly rely on pilot training signals, resulting in significant training overheads (time, energy, spectrum). Digital twin-aided communications have been proposed in \cite{alkhateeb2023real} to reduce or eliminate this overhead by approximating the real world with a digital replica. However, how to implement a digital twin-aided communication system brings new challenges. In particular, how to model the 3D environment and the associated EM properties, as well as how to update the environment dynamics in a coherent manner. To address these challenges, motivated by the latest advancements in computer vision, 3D reconstruction and neural radiance field, we propose an end-to-end deep learning framework for future generation wireless systems that can reconstruct the 3D EM field covered by a wireless access point, based on widely available crowd-sourced world-locked wireless samples between the access point and the devices. This visionary framework is grounded in classical EM theory and employs deep learning models to learn the EM properties and interaction behaviors of the objects in the environment. Simulation results demonstrate that the proposed learnable digital twin can implicitly learn the EM properties of the objects, accurately predict wireless channels, and generalize to changes in the environment, highlighting the prospect of this novel direction for future generation wireless platforms.
\end{abstract}


\vspace{-5pt}

\section{Introduction} \label{sec:Introduction}
Multiple-input multiple-output (MIMO) technology is considered a key feature in 5G and future communication systems \cite{boccardi2014five}. With large-scale antenna arrays, MIMO can provide high multiplexing and array gains, achieving high data rates. However, realizing the full potential of large-scale MIMO systems requires accurate knowledge of the channels. Current communication systems typically rely on training pilot signals to obtain channel information. However, the associated channel acquisition overhead (time, energy as well as spectrum resources) increases proportionally with the number of antennas, hindering MIMO systems from scaling their gains. 

In our recent line of work \cite{alkhateeb2023real, jiang2023digital, jiang2024digitalcsi}, we proposed a novel approach that leverages digital twins to aid wireless communication systems, which can significantly reduce or even eliminate channel acquisition overhead if implemented. The digital twin-aided communication assumes that the base station (BS) has access to a digital replica of the real-world communication environment. This digital replica includes detailed information about the position, orientation, dynamics, shapes, and materials of the communication devices and surrounding objects. By applying electromagnetic (EM) ray tracing within this digital replica, the BS can simulate communication channels. This synthetic channel information can then be used to enhance channel acquisition and guide wireless system operations. With a sufficiently accurate digital replica, conventional channel acquisition can even be bypassed.

While the prior work in \cite{alkhateeb2023real, jiang2023digital, jiang2024digitalcsi} proposed the idea of digital twin-aided communications, it does not discuss how to implement the digital twins. In particular, how can the EM property and interaction behavior of the communication environment be accurately modeled? Moreover, how can the EM 3D model be updated and adapted to changes in the communication environment? We propose learnable digital twins to address these challenges with an end-to-end machine learning (ML) approach.

\subsection{Prior Work}
\textbf{RF Fingerprinting.} In \cite{alrabeiah2019deep}, the author proposed a bijection mapping between the position and the wireless channel in a static environment. Based on the same idea, channel fingerprinting has been widely studied. This research direction considers the wireless channel information as a unique identifier to infer various properties of communication devices, such as device identification \cite{hua2018accurate}, localization \cite{wu2012csi, wang2016csi}, and intrusion detection \cite{jagannath2022comprehensive}. Among these applications, user localization through channel fingerprinting has garnered the most interest. Most prior work in this field employs machine learning to solve the user localization problem because the exact relationship between the wireless channel and user position is challenging to characterize \cite{wu2012csi}, \cite{wang2016csi}, \cite{koike2020fingerprinting}. However, existing work on channel fingerprinting for user localization often assumes fixed communication environments. Additionally, prior models often take a brute-force approach to characterize the function between the wireless channel and user position without explicitly considering electromagnetic (EM) propagation paths and effects. This limits channel fingerprinting approaches to achieve optimal performance and to generalize well in changing environments.

\textbf{Ray Tracing and Propagation Effects.} Ray tracing has been proposed to model the EM wave propagation \cite{yun2015ray}. Compared to the statistical channel models, ray tracing-based channel modeling explicitly tracks propagation paths and models the propagation effects given a 3D EM model, therefore providing deterministic wireless channels that align to the real-world channel at a sample level (instead of the distribution level). Prior work of ray tracing has investigated efficient ways to find all propagation paths (\textit{i.e.} geometric ray tracing) \cite{geok2018comprehensive}. In \cite{deng2017toward}, the author discussed efficient hardware architecture to accelerate ray tracing computation. Efficient spatial data structures to accelerate ray tracing are reviewed in \cite{meister2021survey}. With the development of efficient ray tracing algorithms and hardware, real-time ray tracing is become more and more feasible. 

Prior work has also investigated approaches to model various propagation effects. Calculations of the direct rays and reflected rays are straightforward with the Friis equation, Snell's law, and Fresnel equation. \cite{kouyoumjian1974uniform} proposed the well-known uniform theory of diffraction (UTD), which models the diffraction effect at perfect electric conductors. Heuristic methods, \textit{e.g.}, \cite{holm2000new}, were developed to extend UTD to non-perfectly conducting wedges. In \cite{degli2007measurement}, the author proposed a heuristic mathematical model for the diffuse scattering effects, and validated the proposed model with field measurements.

The existing ray tracing and propagation models are efficient in computation and achieve accurate channel modeling in relatively simple environments. However, these ray tracing channel modeling approaches all assume the knowledge of the physical EM parameters (\textit{conductivity, permittivity, etc.}) of the communication environment. Existing approaches to obtain these physical EM parameters often require specific measurements in controlled environments. Therefore, it is difficult to obtain and update these physical EM parameters in real-world communication sites. Moreover, these physical EM parameters may not fully characterize the EM property of the communication environment, which impacts the accuracy when modeling the propagation effects.

Recent studies have explored the integration of physical-based ray tracing simulators with neural networks \cite{orekondy2023winert, hoydis2023learning}. Specifically, \cite{orekondy2023winert} proposed a neural surrogate to model wireless electromagnetic propagation effects. However, \cite{orekondy2023winert} does not capture more complex interaction effects (\textit{e.g.}, diffraction and diffuse scattering), and requires per-path channel information (\textit{e.g.} gain and delay) as training data that may be expensive to collect in real-world deployments. In contrast, our proposed framework in this paper models complex interaction effects and can be directly trained with point-to-point channel data (without requiring path-specific information). \cite{hoydis2023learning} leveraged neural networks and differentiable ray tracing to learn radio environments, demonstrating promising results. However, their focus on learning and calibrating physical EM properties for ray tracing simulators differs from our work which concentrates on constructing and leveraging wireless digital twins for channel prediction with more generic learning of both EM properties and interaction representations. Further, by decoupling the EM interactions into ray-dependent and ray-independent components, our  framework supports  adaptation to changes and dynamics in the environment (\textit{i.e.}, object movements), which is essential for real-world deployment.

\textbf{Neural Radiance Field.} The neural radiance field (NeRF) \cite{posenc} represents a significant advancement in computer vision and has shown promising performance in novel view synthesis with sparse samples \cite{muller2022instant}.  Specifically, NeRF renders each pixel in the images by accumulating the radiance of a ray along the observing direction. After training, NeRF can synthesize new images from new positions with different viewing directions. In \cite{zhao2023nerf2}, NeRF concept is employed to represent the scene and is further used to represent the wireless propagation. However, it can only model a static environment where the 3D space is discretized into numerous fixed voxels. Any changes in the scene would invalidate the whole neural network. 
However, the use of sparse samples to reconstruct a scene sheds light on how to implement a wireless digital twin! Since in a wireless system, there is always a large amount of wireless samples continuously available. These samples capture valuable information about the environment. This motivates us to think how to exploit them to reconstruct the 3D EM field of the scene.  

\subsection{Contribution}
We propose a new system design paradigm for future wireless platforms—such as self-driving cars, drones, robots, and VR/AR wearables—known as learnable digital twin-aided communications. This innovative approach, which builds upon the concept of \textit{true digital twins} that we introduced for the first time in \cite{alkhateeb2023real}, assumes the availability of a 3D geometry map shared by the wireless access point (BS or WiFi AP) and the devices covered by it. This availability is rooted in the latest advancements in computer vision (neural radiance field, sparse sensing and 3D reconstruction) and is reasonable for future generation communication platforms. Also, every communication platform can reasonably assume it knows the 6DoF pose in the shared 3D map – enabled by the real-time SLAM subsystem. Now, given there are a large amount of wireless channel samples between the access point and the devices with 6DoF tags (\textit{i.e.}, world-locked) in the scene, it is totally possible to reconstruct the EM property of the 3D environment. Once it is trained, the framework can be used to obtain close-to ground-truth channel responses given a device’s location, or vice versa, by knowing the received signal, we can derive the device’s location information. These digital twins can continuously learn and refine their approximations of real-world EM environments with the crowd-sourced wireless samples, enabling it to adapt to environment changes. 
The contribution in this work is summarized below
\begin{itemize}
	\item We proposed an end-to-end deep learning framework to reconstruct the 3D EM field given crowd-sourced wireless samples in a scene. This framework can handle environment changes as well as refine itself over time via crowd-sourced samples in a data-driven manner.
	\item We decoupled the deep learning framework into a ray-independent function and a ray-dependent function. The ray-independent function learns the high-dimensional EM properties of an object in the scene, while the ray-dependent function learns the outgoing E-field given the incoming E-field and the learned EM properties.
	\item We rooted our work in classical EM theory but made generic assumptions to model the complex transformation of different radio wave interaction types (\textit{i.e.}, reflection, diffraction, and scattering). This way, we ensure the generalization of the framework.
	\item We applied this framework to wireless channel estimation, which is a critical part of modern wireless systems, and proved significant performance gain and good generalization to environment changes.
	\item The proposed framework has a wide application scope ranging from UAVs and self-driving cars to AR glass wearables for future-generation wireless systems.
\end{itemize}

To evaluate the proposed learnable digital twin, we build a synthetic dataset of co-existing wireless channels and 3D geometry model. Simulation results demonstrate the efficiency of the proposed learnable digital twin: The learnable digital twin can learn EM property and interaction behaviors of the object in the communication environment. Once trained, the learnable digital twin can achieve high channel prediction accuracy without pilot training overhead. The learnable digital twin can also handle environment changes, which is critical for its real-world application.

The rest of the paper is organized as follows. \sref{sec:dt_comm} introduces the digital twin-aided communications and discusses the key challenges. \sref{sec:learnable_dt} proposes the idea of learnable digital twin, explains the system and channel models, and presents the problem formulation. \sref{sec:Proposed Solution} proposes the ML framework for the learnable digital twin. \sref{sec:end_to_end} presents the end-to-end operation of the training process of the learnable digital twin framework. Experiment setup and results are presented in \sref{sec:experiment_setup} and \sref{sec:result}. \sref{sec:Conclusion} concludes the paper.

\begin{table*}[]
	\caption{Some important variables for problem description in this paper}
	\renewcommand{\arraystretch}{1.5}
	\label{tbl:variable}
	\centering
	\begin{tabular}{cc}
		\toprule
		\multicolumn{2}{c}{Digital Twin Aided Communications}\\
		\midrule
		$\cH$, $g$, $\cE$, $\cD$ & real-world wireless channels, signal propagation law, communication environment, hardware characteristics\\
		$\widetilde{\cH}$, $\widetilde{g}$, $\widetilde{\cE}$ & synthetic channels, ray tracing, EM 3D model\\
		$\widetilde{\mathscr{G}}$, $\widetilde{\mathscr{E}}$ & 3D geometry model, EM property information\\
		$\widetilde{g}_I(\cdot)$, $\widetilde{g}_{\mathscr{G}}$ & ray tracing sub-function that tracks all propagation paths, ray tracing sub-function that models distortion effects\\
		\bottomrule
		\toprule
		\multicolumn{2}{c}{Channel Model}\\
		\midrule
		$N_t$, $N_r$ & number of transmit and receive antennas \\
		$\lambda$, $f$, $\Delta_f$, $K$ & wavelength, carrier frequency, subcarrier spacing, number of subcarriers\\
		$L$ & total number of paths, complex gain, propagation delay, number of interactions of the $l$-th path\\
		$\alpha_l$, $\tau_l$, $I_l$ & complex gain, propagation delay, number of interactions of the $l$-th path\\
		$\theta_l^\mathrm{AoD}$, $\theta_l^\mathrm{AoD}$ & azimuth and elevation AoD (at the transmit antenna) of the $l$-th path\\
		$\varphi_l^\mathrm{AoA}$, $\varphi_l^\mathrm{AoA}$ & azimuth and elevation AoA (at the receive antenna) of the $l$-th path\\
		$\bc_\mathrm{tx}(\theta, \varphi), \bc_\mathrm{rx}(\theta, \varphi)$ & antenna patterns of the transmit and receive antennas\\
		$T_l(\cdot)$ & transfer function of the $l$-th path\\
		$T_{l,i}(\cdot)$, $\bT_{l,i}$ & linear transfer matrix or nonlinear transfer function at the $i$-th interaction of the $l$-th path\\
		$\bE_{l,i}$ & incoming E-field at the $i$-th interaction of the $l$-th path\\
		\bottomrule
		\toprule
		\multicolumn{2}{c}{Learnable Digital Twin Framework}\\
		\midrule
		$\mathscr{G}_o$ & 3D geometry model of the $o$-th object\\
		$\cF_o$, $\cB_o$, $\cV_o$ & surfaces, edges, and vertices in the 3D geometry model of the $o$-th object\\
		$\bp_o$, $\br_o$& position and orientation of the 3D geometry model of the $o$-th object\\
		$\cG_{l,i}$ & geometry path parameters at the $i$-th interaction of the $l$-th path\\
		$\bp_{l,i}$, $o_{l,i}$, $\cI_{l,i}$& interaction position, object index, and geometry information at the $i$-th interaction of the $l$-th path\\
		$\bd_{l,i}^\mathrm{AoA}$ , $\bd_{l,i}^\mathrm{AoD}$ & AoA and AoD directions at the $i$-th interaction of the $l$-th path\\
		$\widetilde{g}_{\mathscr{E}}(\mathbf{\Theta}_\mathscr{E})$ & DL model that encodes the EM property of the communication environment\\
		$\widetilde{g}_{T}(\mathbf{\Theta}_T)$ & DL model that predicts the transfer function\\
		\bottomrule
	\end{tabular}
\end{table*}

\section{Digital Twin Aided Communications}\label{sec:dt_comm}
In \cite{alkhateeb2023real, jiang2023digital, jiang2024digitalcsi}, we proposed a novel direction of real-time digital twin-aided communications. By approximating the real world with a digital replica, the communication system can infer/predict the real-world channel information with eliminated (or significantly reduced) channel acquisition overhead. The digital twin can be leveraged to aid the communication systems in mainly two ways: (i) The predicted channel information/parameters can be leveraged to make (near) real-time and proactive decisions for the communication systems, and (ii) the digital replica can be used to build high-fidelity and site-specific synthetic datasets that could be utilized to train machine learning models for various wireless communication tasks. Next, we introduce the components and key challenges of digital twins.

\begin{figure*}[t]
	\centering
	\includegraphics[width=1.\linewidth]{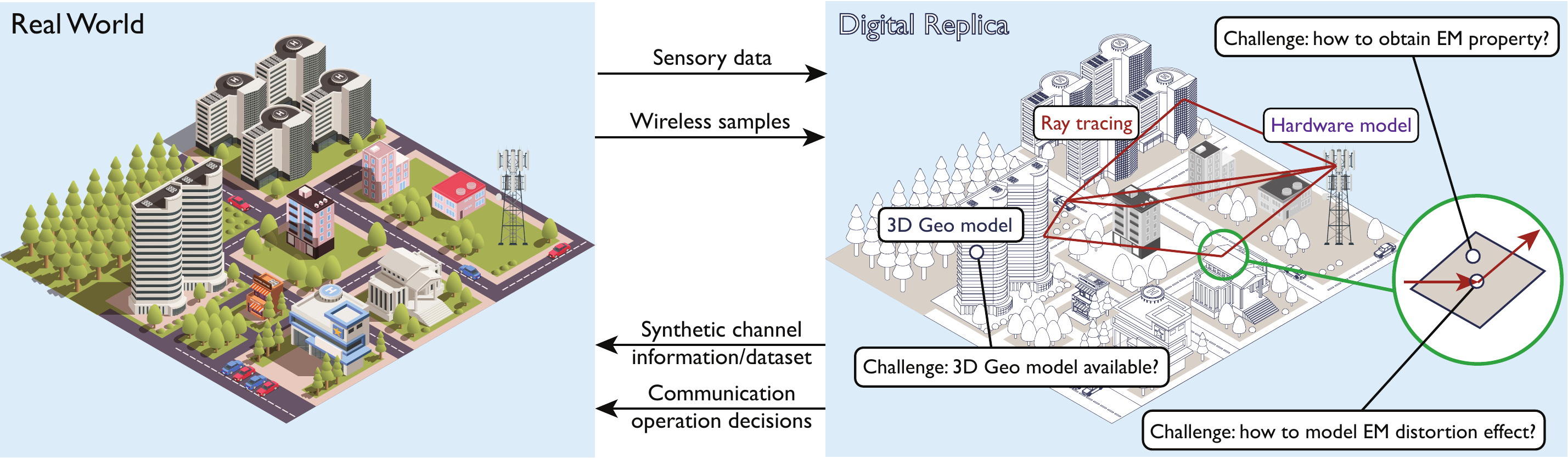}
	\caption{This figure captures the components of the digital replica, key challenges in digital twin construction, and the applications of digital-twin aided communications. In particular, the key challenges in digital twin construction includes: (i) how to obtain geometry information, (ii) how to obtain EM property information, and (iii) how to model EM distortion effect.}
	\label{fig:key_challenges}
\end{figure*}

\subsection{Digital Replica Components}
The real-world communication channel is determined by: (i) the communication environment denoted by $\cE$, which includes the positions, orientations, dynamics, shapes, and EM properties of the BS, the UEs, and the other surrounding objects (scatterers), (ii) the wireless signal propagation law denoted by $g(\cdot)$, and (iii) the hardware characteristics (\textit{e.g.} antenna patterns) of the communication devices (BSs and UEs) denoted by $\cD$. Given the communication environment, wireless signal propagation law, and the device hardware characteristics, the communication channels $\cH$ can be written as
\begin{align}\label{eq:real_world}
	\cH = g(\cE, \cD).
\end{align}
To approximate real-world communication channels, digital twins approximate the communication environment $\cE$ with EM 3D models and the signal propagation law $g(\cdot)$ with EM simulations (\textit{e.g.}, ray-tracing). Next, we elaborate more on the EM 3D model, ray-tracing, and hardware model. 

\textbf{EM 3D model.} The EM 3D model ${\cE}$ contains the (estimated) information about the positions, orientations, dynamics, shapes, and EM properties of the BS, the UE, and other surrounding objects (reflectors/scatterers) in the communication environment. We further divide the EM 3D model into two components $\widetilde{\cE} = (\widetilde{\mathscr{G}}, \widetilde{\mathscr{E}})$. $\widetilde{\mathscr{G}}$ denotes the 3D \textit{geometry} model that contains the geometric information of the EM 3D model, including the position, orientation, dynamic, and shape. The other component, $\widetilde{\mathscr{E}}$, contains the \textit{EM property} information of the EM 3D model.

\textbf{Ray tracing.} The ray tracing, denoted as $\widetilde{g}(\cdot)$, simulates the propagation paths between each transmit-receive antenna pair. This simulation utilizes the geometry and EM property information from the EM 3D models and accounts for multiple propagation effects, \textit{e.g.}, reflection, diffraction, and scattering. For each path, ray tracing produces the path parameters such as the complex path gain, propagation delay, and propagation angles. These path parameters can then be used to synthesize communication channels. We further decouple the ray tracing into two functions: (i) $\widetilde{g}_{\mathscr{G}}(\widetilde{\mathscr{G}})$, which tracks all the propagation paths between the transmit-receive antenna pair, and (ii) $\widetilde{g}_{I}\big(\widetilde{g}_{\mathscr{G}}(\widetilde{\mathscr{G}}), \widetilde{\mathscr{E}}, \cD \big)$, which models the distortion (\textit{i.e.}, changes) of the EM wave along propagation paths, and synthesizes the channel\footnote{$\widetilde{g}_{\mathscr{G}}(\widetilde{\mathscr{G}})$ outputs the propagation delay and angle of all possible candidate paths based solely on the 3D geometry model. $\widetilde{g}_{I}\big(\widetilde{g}_{\mathscr{G}}(\widetilde{\mathscr{G}}), \widetilde{\mathscr{E}}, \cD\big)$ may eliminate some propagation paths considering the EM properties and hardware characteristics by assigning zero (or near-zero) path gain.}.

\textbf{Hardware model.} The hardware model encapsulates the characteristics of the communication hardware, ranging from the antenna pattern to the baseband analog-to-digital converter (ADC). Ideally, this model should accurately represent characteristics of the hardware transceiver and antenna arrays. In this paper, we assume that the hardware characteristics $\cD$ can be measured in controlled environments by the device manufacturer/vendor, and the hardware model is available at the BS.

The digital replica can be used to generate synthetic communication channels, which can be written as
\begin{align}\label{eq:dt}
	\widetilde{\cH} &= \widetilde{g}\Big(\widetilde{\cE}\Big) \nonumber\\
	&=\widetilde{g}_{I}\bigg(\widetilde{g}_{\mathscr{G}}\Big(\widetilde{\mathscr{G}}\Big), \widetilde{\mathscr{E}}, \widetilde{\mathscr{G}}, \cD\bigg).
\end{align}
With accurate EM 3D model, ray tracing, and hardware model, the synthetic channels $\widetilde{\cH}$ could potentially be a close approximation of the real-world channels $\cH$. With that, the generated synthetic channel information can be used to aid various wireless communication operations.

\subsection{Digital Twin Key Challenges}\label{sec:key_challenge}
Here, we analyze \eqref{eq:dt} and present the key challenges of constructing the digital replica. Observing \eqref{eq:dt}, we can see that some components of the digital replica are relatively easy to solve. In particular, the hardware characteristics is fixed given a communication device. The shooting-and-bouncing ray (SBR) method \cite{yun2015ray} can be employed in $\widetilde{g}_{\mathscr{G}}(\widetilde{\mathscr{G}})$ to track the propagation paths and obtain the corresponding propagation delay and angles. However, constructing the digital replica incorporates the following key changes.
\begin{itemize}
	\item $\widetilde{\mathscr{G}}$: The digital replica requires near real-time geometry information about the surrounding communication environment to construct the geometry 3D model $\widetilde{\mathscr{G}}$. This real-time geometry information is typically not available in today’s conventional wireless communication systems.
	\item $\widetilde{\mathscr{E}}$: Obtaining accurate information about the EM property $\widetilde{\mathscr{E}}$ of the communication environment is challenging for the following reasons: (i) traditional approaches can measure the physical EM parameters (\textit{e.g.} conductivity) of an object. However, this measurement often requires specific devices and controlled environments. Therefore, measuring all objects in real-world communication sites using the traditional approaches is not practical, (ii) the commonly considered (fixed) physical EM parameters, such as the conductivity and the permittivity, may not always be sufficient to fully characterize the EM properties. For instance, the EM properties can be anisotropic and dependent to environmental conditions such as temperature and humidity, and (iii) the communication environment may incorporate non-uniform materials whose EM properties are difficult to measure and characterize.
	\item $\widetilde{g}_{I}(\cdot)$: The $\widetilde{g}_{I}(\cdot)$ models the distortion effects of the EM wave along propagation. For straightforward scenarios (\textit{e.g.}, reflection of an infinitely large, uniform, non-magnetic surface), the distortion effects can be obtained directly from Fresnel equation. However, deriving closed-form solutions for more complex distortions, such as diffraction and diffuse scattering, proves challenging. The task of modeling distortion effects becomes even more difficult when considering scenarios where the reflectors or scatters exhibit non-uniform EM properties or have more intricate shapes.
\end{itemize}
In the next section, we propose a learnable digital twin framework and attempt to address these challenges in constructing the digital replica.

\section{Learnable Digital Twin Framework}\label{sec:learnable_dt}
In this section, we first introduce the system model for the learnable digital twin-aided communications. Then, we present the key concept of the learnable digital twin.

\subsection{System Model}\label{sec:system model}
We consider a future-generation MIMO communication system where a BS, which can be a cellular BS or WiFi access point (AP), serves $M$ UE devices. The BS is equipped with an antenna array consisting of $N_t$ elements, and the UE features an antenna array with $N_r$ elements. The system employs the OFDM modulation with $K$ subcarriers for the communication signals. Note that while we consider the OFDM modulation in this paper, the proposed learnable digital twin can be straightforwardly generalized to other modulations and waveforms.

The following three key features distinguish future-generation communication systems from current conventional systems.
\begin{itemize}
	\item Access to 3D geometry map: Consider the 3D geometry of the communication environment covered by the MIMO system, it is composed of the static structures of the scene and the objects therein. Recent computer vision advancements present many methods to reconstruct the 3D geometry and segment the objects \cite{nie2020total3dunderstanding, zhang2021holistic, zhang2021deeppanocontext}. Recent research \cite{qi_link3} proposed a novel method for reconstructing environments and representing the layout of physical spaces while being able to describe objects, states, and complex geometry. Therefore, considering the always-on sensing capability that will equip with future communication platforms, it is totally feasible to reconstruct the 3D geometry environment, and to assume access to a shared 3D geometry map of their environment in the future.
	\item Access to real-time 6DoF pose information: We can assume the user devices (\textit{e.g.}, wearable AR glass, self-driving car, drone, robot) have a real-time SLAM subsystem that provides accurate 6DoF pose in the 3D geometry map.
	\item Access to world-locked crowd-sourced wireless samples: A large population of user devices can continuously provide crowd-sourced wireless (channel) samples, each tagged with precise 6DoF positions.
\end{itemize}
These features enable our vision for the learnable digital twin and raise opportunities for cross-discipline design and optimization. We base our work here on these assumptions.

\subsection{Key Idea}
Assuming the communication system has access to the 3D geometry model of the communication environment, two critical challenges remain in constructing the digital twin: (i) obtaining the EM property $\widetilde{\mathscr{E}}$ of the communication environment and (ii) modeling the interaction behavior (distortion effect) of EM propagation $\widetilde{g}_I(\cdot)$. To that end, we propose to leverage the communication system to capture information about the EM property and the distortion effect, and learn the EM property and the distortion effect directly from communication channels\footnote{It may also be possible to learn the EM property and the distortion effect from other wireless data such as the received time domain pilot waveform. We leverage the communication channels because it is directly available at most communication devices.} in an end-to-end manner.

Let $\widetilde{g}_{\mathscr{E}}({\mathbf{\Theta}}_\mathscr{E})$ denote a DL model (neural field) that encodes the EM property of the communication environment, with ${\mathbf{\Theta}}_\mathscr{E}$ representing the trainable model parameters. Let $\widetilde{g}_I({{\mathbf{\Theta}}_I})$ denote a DL model that describes the interaction behavior. The $\widetilde{g}_I({{\mathbf{\Theta}}_I})$ can be used to predict the wireless channels based on the encoded EM property in $\widetilde{g}_{\mathscr{E}}({\mathbf{\Theta}}_\mathscr{E})$ as follows
\begin{align}
	\widetilde{\cH}=\widetilde{g}_{I}\bigg(\widetilde{g}_{\mathscr{G}}\Big(\widetilde{\mathscr{G}}\Big), \widetilde{g}_{\mathscr{E}}({\mathbf{\Theta}}_\mathscr{E}), \mathscr{G}, \cD; \mathbf{\Theta}_I \bigg).
\end{align}
The objective of the learnable digital twin can then be written as
\begin{align}\label{eq:dt_abstract}
	\min_{\widetilde{g}_{\mathscr{E}}({\mathbf{\Theta}}_\mathscr{E}),\ \widetilde{g}_I({{\mathbf{\Theta}}_I})}\bbE_{\cH\sim \cE}\bigg[J\big(\cH, \widetilde{\cH}\big)\bigg],
\end{align}
where $\cE$ is a communication site, and $\cH$ is a channel sampled from the communication site $\cE$. $J(\cH, \widetilde{\cH})$ denotes the discrepancy between real-world and synthetic channels.

\begin{figure}[t]
	\centering
	\includegraphics[width=1.\linewidth]{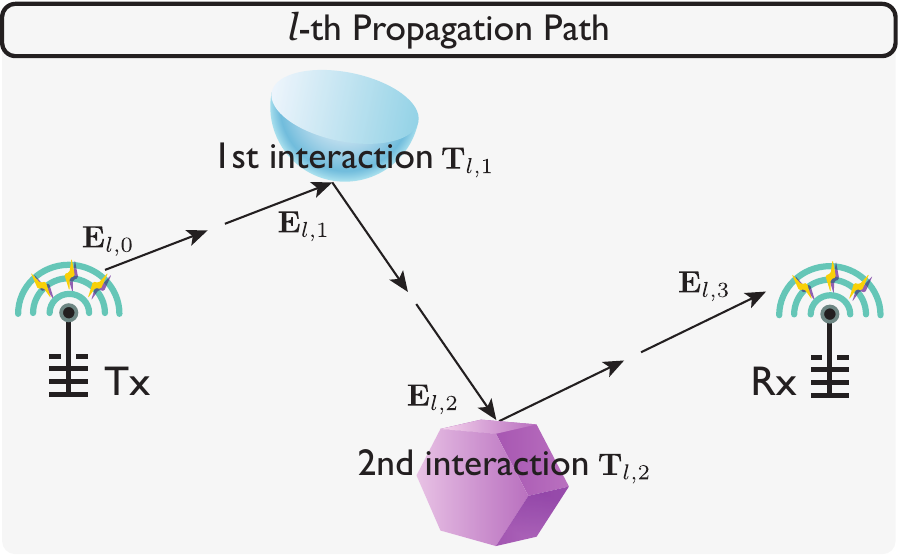}
	\caption{This figure illustrates the $l$-th propagation path. The electric field emitted by the transmit antenna goes through two interactions and then reaches the receive antenna.}
	\label{fig:interaction}
\end{figure}

\section{Channel Model and Transfer Function}\label{sec:transfer_function}
In the previous section, we present the key idea of the learnable digital twin. To further concertize \eqref{eq:dt_abstract}, we elaborate on the channel model and transfer functions for EM interactions.

\subsection{Channel Model}
We consider a multi-path block-fading wideband geometric channel model. Let \mbox{$h_{n_r, n_t}(f)\in\bbC$} denote the channel frequency response (CFR) between the $n_t$-th BS antenna and the $n_r$-th user antenna at frequency $f$. The CFR $h_{n_r, n_t}(f)$ can be written as
\begin{align}\label{eq:cfr}
	h_{n_r, n_t}(f) = \sum_{l=1}^L \alpha_l e^{-j2\pi f\tau_l},
\end{align}
where $L$ is the number of propagation paths, and $\alpha_l\in\bbC$ and $\tau_l\in\bbR$ are the complex amplitude gain and propagation delay of the $l$-th path ($l=\{1,\hdots,L\}$). Let $f_c$ and $\Delta_f$ denote the carrier frequency and the subcarrier spacing of the OFDM system, the frequency-domain OFDM channel between the $n_t$-th BS antenna and $n_r$-th UE antenna $\bh_{n_r,n_t}\in\bbC^{K\times 1}$ is then given by
\begin{align}\label{eq:2}
	\bh_{n_r,n_t} = \big[ h_{n_r, n_t}[0], h_{n_r, n_t}[1], \hdots, h_{n_r, n_t}[K-1]\big]^T,
\end{align}
where $h_{n_r,n_t}[k] = h_{n_r, n_t}(f_c + (k-\frac{K}{2})\Delta_f)$ is the frequency response on the $k$-th subcarrier. The frequency-domain OFDM channel between the BS and the UE $\bH\in\bbC^{N_rN_t\times K}$ can then be written as
\begin{align}\label{eq:2_5}
	\bH = \big[ \bh_{1, 1}, \bh_{1, 2}, \hdots, \bh_{1, N_t}, \hdots, \bh_{N_r, N_t}\big]^T.
\end{align}

The complex amplitude gain $\alpha_l$ in \eqref{eq:cfr} depends on the EM distortion effect along the $l$-th propagation path and the hardware characteristics, which can be written as \cite{fugen2006capability}
\begin{align}\label{eq:3}
	\alpha_l = \frac{\lambda}{4 \pi} a_l\bc_\mathrm{rx}^H\left(\theta_l^{\mathrm{AoA}}, \varphi_l^{\mathrm{AoA}}\right) T_l\bigg(\bc_{\mathrm{tx}}\left(\theta_l^{\mathrm{AoD}}, \varphi_l^{\mathrm{AoD}}\right)\bigg),
\end{align}
where $\lambda$ is the wavelength. $\theta_l^{\mathrm{AoA}}$ and $\varphi_l^{\mathrm{AoA}}$ are the elevation and azimuth angle of arrival (AoA) of the $l$-th path, respectively. $\bc_\mathrm{rx}\left(\theta_l^{\mathrm{AoA}}, \varphi_l^{\mathrm{AoA}}\right) \in \bbC^{2\times 1}$ is the receive (UE) antenna's amplitude gain of the two orthogonal polarization directions in the spherical coordinate system, parameterized by the two AoAs. $\theta_l^{\mathrm{AoD}}$, $\varphi_l^{\mathrm{AoD}}$, and $\bc_{\mathrm{tx}}$ are similarly defined for the transmitter (BS) side. Note that here, we only consider the antenna patterns for simplicity. The hardware characteristics of other components along the analog processing chain may also need to be considered. $a_l \in \bbR$ is the spread factor associated with the geometrical spreading out of the propagation path. $a_l$ depends on the wavefront and the geometry of the propagation (\textit{e.g.}, propagation distances) \cite{pathak2013uniform, paknys2016uniform}. $T_l\colon \bbC^{2 \times 1} \to \bbC^{2 \times 1}$ is the transfer function of the $l$-th path, which accounts for the distortion effects (with $a_l$ factored out) along this propagation path. Note that $\alpha_l$ directly combines the two polarization components with the $2$-by-$1$ receive antenna pattern $\bc_\mathrm{rx}$.

\subsection{Transfer Function}
As shown in \figref{fig:interaction}, while propagating from the transmit antenna to the receive antenna, the electric field (E-field) can be distorted by multiple objects along a propagation path. We define an \textit{interaction} as any event where an object along the signal propagation path causes distortion or alteration of the E-field. An E-field can be written as
\begin{align}
	\bE = E_s \hat{\bee}_s + E_p \hat{\bee}_p,
\end{align}
where $\hat{\bee}_s, \hat{\bee}_p \in \bbC^{3\times 1}$ are unit vectors denoting the two orthogonal polarization directions. $E_s, E_p \in \bbC$ are the complex amplitude of the E-field on these polarization directions. Let $\bE_{l, i}$ denote the incoming E-field at the $i$-th interaction, the incoming (incident) E-field $\bE_{l, i+1}$ at the $(i+1)$-th interaction (or the receive antenna) can then be written as
\begin{align} \label{eq:one_interaction}
	\bE_{l, i+1} = T_{l, i}\left(\bE_{l, i}\right),
\end{align}
where $T_{l, i}\colon \bbC^{2\times 1} \to \bbC^{2\times 1}$ denote the distortion effect (transfer function) of the $i$-th interaction along the $l$-th path. Note that $T_{l, i}(\cdot)$ and $\bE_{l,i}$ do not account for the phase change due to the propagation delay because this phase term has been included (addressed) in \eqref{eq:cfr}. Specifically, we define
\begin{subequations} \label{eq:first_interaction}
	\begin{align} 
		\bE_{l, 0} &= \bc_{\mathrm{tx}}\left(\theta_l^{\mathrm{AoD}}, \varphi_l^{\mathrm{AoD}}\right)\\
		\bE_{l, 1} &= T_{l, 0}\left(\bE_{l, 0}\right) = \frac{\bE_{l, 0}}{d_{l, 1}},
	\end{align}
\end{subequations}
where $d_{l, 1}$ is the distance between the transmit antenna and the first interaction position (or the receive antenna if the path is line-of-sight). Note that the $\frac{1}{d_{l, 1}}$ term accounts for the free-pace pathloss from the transmit antenna to the first interaction point.

Combining \eqref{eq:one_interaction} and \eqref{eq:first_interaction}, the relationship between the E-filed emitted by the transmit antenna and the one that reaches the receive antenna can be written as
\begin{align} \label{eq:interaction}
	\bE_{l, I_l} &= T_{l, I_l}\Bigg(\hdots \bigg( T_{l, 1} \Big(T_{l, 0} \big(\bE_{l, 0}\big) \Big) \bigg)\Bigg) \nonumber\\
	&= T_l\Big( \bc_{\mathrm{tx}}\big(\theta_l^{\mathrm{AoD}}, \varphi_l^{\mathrm{AoD}}\big)\Big),
\end{align}
where $I_l$ is the total number of interactions along the $l$-th path. The transfer function of the $l$-th path $T_{l}$ is the cascadation of the transfer functions of all interactions along the path $T_{l,0} \hdots, T_{l,I_l}$. The transfer function $T_{l,i}$ depends on the interaction type (\textit{e.g.} reflection, diffraction, etc.), EM property of the interaction point, and geometric parameters of the interaction (\textit{e.g.} incoming angle of the EM wave). If we consider a simple yet important case where the communication channel and the transfer functions are linear, the transfer function $T_{l,i}\colon \bbC^{2 \times 1} \to \bbC^{2 \times 1}$ reduces to a linear transfer matrix $\bT_{l,i}\in\bbC^{2\times2}$ such that $T_{l,i}(\bE) = \bT_{l,i}\bE$. With that, \eqref{eq:interaction} can be re-written as
\begin{align}
	\bE_{l, I_l} &= \bigg( \prod_{i=0}^{I_l} \bT_{l, i}\bigg) \bc_{\mathrm{tx}}\big(\theta_l^{\mathrm{AoD}}, \varphi_l^{\mathrm{AoD}}\big) \nonumber\\
	&= \bT_{l} \bc_{\mathrm{tx}}\big(\theta_l^{\mathrm{AoD}}, \varphi_l^{\mathrm{AoD}}\big),
\end{align}
where $\bT_l \in \bbC^{2\times 2}$ is the linear transfer matrix corresponding to the $l$-th path. In this paper, we consider both the linear and nonlinear transfer functions.

\section{Problem Formulation}
By applying ray tracing $\widetilde{g}_\mathscr{G}$ to the geometry 3D model of the communication environment $\widetilde{\mathscr{G}}$, we can track the propagation paths and obtain the prorogation delay and angles $\tau_l,\theta_l^\mathrm{AoD},\varphi_l^\mathrm{AoD},\theta_l^\mathrm{AoA}, \varphi_l^\mathrm{AoA}\ \forall l=\{1, \hdots, L\}$. Observing \eqref{eq:cfr} and \eqref{eq:3}, we see that the unknown term to synthesize the communication channel is $T_l,\ \forall l=\{1, \hdots, L\}$. That is, the learnable digital twin can be achieved by learning the transfer functions of the propagation paths. Let $\widetilde{g}_{T}({\mathbf{\Theta}}_T)$ denote the DL model that predicts the transfer function of one interaction, with $\mathbf{\Theta}_T$ representing the trainable parameters. Let $\bE_{l, i}$ denote the incoming E-field of the interaction, the outgoing E-field can then be represented as
\begin{align}\label{eq:prob_form}
	\bE_{l, i+1} = \widetilde{g}_{T}\bigg(\widetilde{g}_{\mathscr{G}}(\widetilde{\mathscr{G}}), \widetilde{g}_{\mathscr{E}}(\mathbf{\Theta}_\mathscr{E}),\bE_{l, i} ;\mathbf{\Theta}_T\bigg).
\end{align}
Substituting \eqref{eq:3}, \eqref{eq:one_interaction}, \eqref{eq:interaction} and \eqref{eq:prob_form} into \eqref{eq:cfr}, we can obtain the wireless channel predicted by the learnable digital twin $\widetilde{\bH}$. The EM property and the interaction effects can then be learned by solving the following optimization problem
\begin{align}\label{eq:dt_abstract}
	\widetilde{g}^\star_{\mathscr{E}}({\mathbf{\Theta}}^\star_\mathscr{E}),\ \widetilde{g}^\star_T({{\mathbf{\Theta}}^\star_T}) = \underset{\widetilde{g}_{\mathscr{E}}({\mathbf{\Theta}}_\mathscr{E}),\ \widetilde{g}_T({{\mathbf{\Theta}}_T})}{\arg\min}\bbE_{\bH\sim \cE}\bigg[J\big(\bH, \widetilde{\bH}\big)\bigg].
\end{align}
In practice, the expected value can be approximated by averaging over the dataset.

\section{Proposed Digital Twin Learning Framework}\label{sec:Proposed Solution}
In this section, we propose a learnable digital twin framework that (i) leverages the 3D geometry model in future system platforms, (ii) applies geometric ray tracing on the 3D geometry model to trace the propagation paths and obtain the geometric path parameters, and (iii) leverage DL models to learn the EM property and interaction behavior of the objects in the communication environment in a data-driven manner. The learned EM property and interaction behavior can then be used to predict the communication channels in the digital replica. Note that the proposed ML framework is designed to handle changes in the scenario and can even be extended to handle dynamic scenarios. However, for simplicity, we explain the proposed learnable digital twin framework assuming a static communication environment. 
\subsection{Real-Time 3D Geometry Model Construction}
As discussed in \sref{sec:system model}, we assume the proposed future-generation communication system can reconstruct the 3D geometry model of the communication environment, and segment, identify, and track the objects therein via advanced computer vision techniques. Let $\cO_o$, ($o=1,\hdots, O$), denote the $o$-th identified object, the 3D geometry model of this object can then be written as
\begin{align}
	\widetilde{\mathscr{G}}_o = (\cF_o, \cB_o, \cV_o, \bp_o, \br_o),
\end{align}
where $\cF_o$, $\cB_o$, and $\cV_o$ denote the identified sets of surfaces, edges, and vertices that belong to the $o$-th object. $\bp_o\in\bbR^{3\times1}$ and $\br_o\in\bbR^{3\times1}$ denote the 3D position and rotation of the $o$-th object. The 3D geometry model of the scene incorporating all $O$ objects can then be written as
\begin{align}
	\widetilde{\mathscr{G}} = \{\widetilde{\mathscr{G}}_{\mathrm{tx}}, \widetilde{\mathscr{G}}_{\mathrm{rx}}, \widetilde{\mathscr{G}}_1, \hdots, \widetilde{\mathscr{G}}_O\},
\end{align}
where $\widetilde{\mathscr{G}}_{\mathrm{tx}}=(\bp_{\mathrm{tx}}, \br_{\mathrm{tx}})$ and $\widetilde{\mathscr{G}}_{\mathrm{rx}}=(\bp_{\mathrm{rx}}, \br_{\mathrm{rx}})$ contain the position and orientation of the transmit and receive antennas, respectively. This paper primarily focuses on learning the EM property and interaction behavior. Therefore, we assume that the 3D geometry model can be perfectly reconstructed.

\subsection{Geometric Ray Tracing}
We apply the shooting-and-bouncing (SBR) method $\widetilde{g}_{\mathscr{G}}(\cdot)$ on the 3D geometry model $\widetilde{\bS}$, which provides the geometric propagation path parameters. In particular, the SBR method begins by sampling initial ray shooting directions on the unit sphere. Then, SBR shoots initial rays from the transmit antenna along the sampled initial directions. For each ray shot into the scene, the SBR algorithm calculates intersections between the ray and the objects. In particular, this step determines which object a ray will hit based on the geometric 3D model, and provides the index of the interaction object, the position of the interaction point, the orientation of the hitting surfaces and edges, and the incoming direction of the ray. Upon hitting an object, the ray can be reflected, diffracted, and diffusely scattered. In the case of reflection, there is only one outgoing ray whose direction follows the law of reflection. In the case of diffraction and diffuse scattering, the SBR again samples multiple outgoing directions starting from the interaction position. The SBR recursively traces the propagation paths and finds the paths that can reach the receive antenna within a predefined depth (\textit{i.e.}, maximum number of interactions). Note that this SBR method only depends on the 3D geometry model, and it does not require the EM property of the objects or model the EM interaction effect.

\begin{figure}[t]
	\centering
	\includegraphics[width=1.\linewidth]{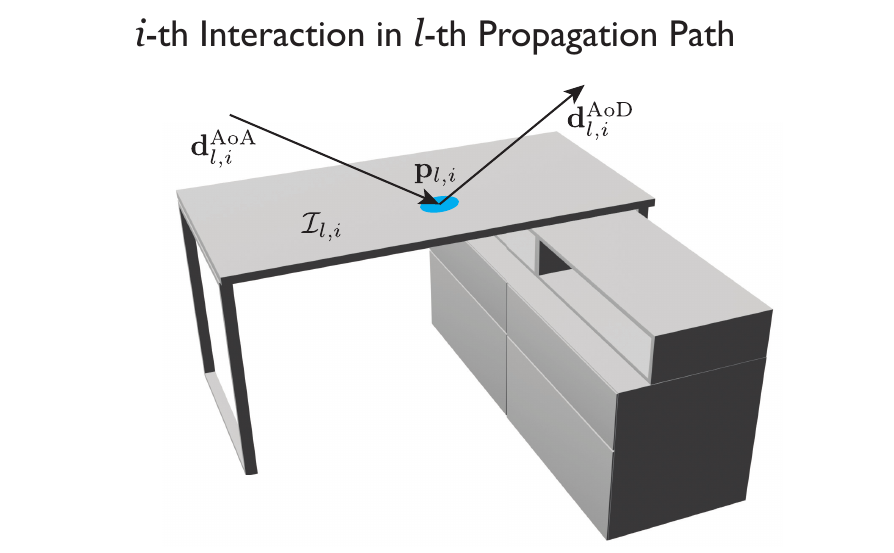}
	\caption{This figure shows the $i$-th interaction at position $\mathbf{p}_{l, i}$ along the $l$-th propagation path. The $\mathbf{d}_{l, i}^{\mathrm{AoA}}, \mathbf{d}_{l, i}^{\mathrm{AoD}}$ are unit vectors representing the direction of the incoming and outgoing rays, respectively. $\cI_{l, i}$ contains the interaction geometry information.}
	\label{fig:one_interaction}
\end{figure}

With the SBR method, we obtain the geometric propagation path parameters of all $L$ paths. As shown in \figref{fig:one_interaction}, consider the $i$-th interaction of the $l$-th path, the geometric propagation path parameters can be written as
\begin{align}\label{eq:geo_path_param}
	\cG_{l, i} = (\bp_{l,i}, o_{l,i}, \mathbf{d}_{l, i}^{\mathrm{AoA}}, \mathbf{d}_{l, i}^{\mathrm{AoD}}, \cI_{l,i}),
\end{align}
where $\mathbf{p}_{l, i}\in \bbR^{3\times1}$ and $o_{l,i}$ denote the interaction position and the index of the interaction object, respectively. $\mathbf{d}_{l, i}^{\mathrm{AoA}}, \mathbf{d}_{l, i}^{\mathrm{AoD}}\in \bbR^{3\times1}$ are unit vectors representing the incoming and outgoing ray directions. $\cI_{l,i}$ contains information about the interaction geometry, which depends on the type of interaction. For reflection and diffuse scattering, we assume that the ray interacts with a surface. In this case, $\cI_{l,i}$ can be written as
\begin{align}
	\cI_{l,i}^{\{\mathrm{refl},\ \mathrm{scat}\}} = (\bn_{l,i}, \kappa_{l,i}),
\end{align}
where $\bn_{l,i}\in \bbR^{3\times1}$ is a unit vector representing the normal of the surface. $\kappa_{l,i}$ indicates the interaction type. For diffraction, the ray interacts with a wedge. In this case, $\cI_{l,i}$ can be written as
\begin{align}
	\cI_{l,i}^\mathrm{diffr} = (\bn^0_{l,i}, \bn^n_{l,i}, \kappa_{l,i}),
\end{align}
where $\bn^0_{l,i}, \bn^n_{l,i}\in \bbR^{3\times1}$ are unit vectors representing the normals of the two surfaces that form the wedge. The final output of the geometry ray tracing is the geometric propagation path parameters between the transmit and receive antenna, which can be written as
\begin{align}\label{eq:ray_tracing_final}
	\cG = \{ \cG_1, \hdots, \cG_L\},
\end{align}
where $L$ is the total number of paths. $\cG_l$ contains the geometric propagation path parameters of all $I_l$ interactions along the $l$-th path, which is given by
\begin{align}
	\cG_l = \{ \cG_{l, 1}, \hdots, \cG_{l,I_l}, \theta_l^{\mathrm{AoD}}, \phi_l^{\mathrm{AoD}}, \theta_l^{\mathrm{AoA}}, \phi_l^{\mathrm{AoA}}, \tau_l\},
\end{align}
where $\theta_l^{\mathrm{AoD}}, \phi_l^{\mathrm{AoD}}$ are the AoDs from the transmit antenna, and $\theta_l^{\mathrm{AoA}}, \phi_l^{\mathrm{AoA}}$ are the AoAs at the receive antenna. Note that the propagation delay $\tau_l$ of the $l$-th path can be easily obtained from the positions of the transmit and receive antennas and the positions of all interaction points along the path.

\subsection{Machine Learning Modeling}\label{sec:ml_modeling}
We propose a novel ML architecture that models the EM property and the interaction behavior. The proposed ML architecture incorporates two components: (i) neural object and (ii) neural interaction. The proposed ML architecture can be trained to learn the EM property and the interaction behavior in a data-driven and end-to-end manner. Once trained, the ML models can then be used to synthesize wireless channels in the digital replica. Next, we elaborate on these two components.
\subsubsection{Neural Object} 
Drawing inspiration from the concept of the neural radiance field (NeRF), we propose to leverage a neural network $\widetilde{g}_\mathscr{E}(\mathbf{\Theta}_\mathscr{E})$ to encapsulate the EM property of the communication environment in \eqref{eq:prob_form}. Given the object segmentation and identification in the geometry 3D model, we propose to further divide the $\widetilde{g}_\mathscr{E}(\mathbf{\Theta}_\mathscr{E})$ into $O$ individual neural network models:
\begin{align}
	\widetilde{g}_\mathscr{E}(\mathbf{\Theta}_\mathscr{E}) = \bigg\{\widetilde{g}_{\mathscr{E}, o}(\mathbf{\Theta}_{\mathscr{E}, o}) \mathrel{\Big|} o=\{1,\hdots, O\} \bigg\},
\end{align}
where $\widetilde{g}_{\mathscr{E}, o}(\mathbf{\Theta}_{\mathscr{E}, o})$ is what we call a ``neural object" because it is used to encode the EM property of one object in the geometry 3D model.

It is important to emphasize that the EM property of an object is intrinsic and determined by its material and shape. That is, for a given object, the EM property can be solely defined by a position in the object's \textit{local coordinates} (assuming fixed EM frequency), and is not dependent on the attributes of the incoming/outgoing E-field (\textit{e.g.} AoA and AoD) and the type of interaction. Let $\bp\in\bbR^{3 \times1}$ denote a position in the \textit{local coordinate} of the $o$-th object, the corresponding high-dimensional encoded representation is given by
\begin{align}
	\bee = \widetilde{g}_{\mathscr{E}, o}\big(\mathrm{PosEnc}(\bar{\bp});\mathbf{\Theta}_{\mathscr{E}, o}\big),
\end{align}
where $\widetilde{g}_{\mathscr{E}, o}(\mathbf{\Theta}_{\mathscr{E}, o})$ is the $o$-th neural object (neural network for the $o$-th object) with $\mathbf{\Theta}_{\mathscr{E}, o}$ representing the trainable parameters. Note that the $\bee$ is a high dim representation of EM property, which does not have a concrete physical meaning. $\bar{\bp}$ is obtained by normalizing $\bp$ with the maximum dimension of the object such that the coordinates in $\bar{\bp}$ lie in [-1,1]. Instead of having the NN operate directly on the input position coordinates $\bp$, we adopt the positional encoding \cite{posenc} given by \eqref{eq:PosEnc}.
\begin{figure*}
	\begin{equation}
		\mathrm{PosEnc}(x) = \big[\sin(2^0\pi x), \cos(2^0\pi x), \hdots, \sin(2^{M-1}\pi x), \cos(2^{M-1}\pi x)\big]^T. \label{eq:PosEnc}
	\end{equation}
\end{figure*}
Note that when the input is a vector, the positional encoding $\mathrm{PosEnc}(\cdot)$ is applied individually to each dimension of the input, and the encoded outputs are concatenated to one output vector. It has been shown that positional encoding is effective for neural networks to represent high-frequency functions. The parameter $M$ controls how fast the EM property can vary across space. In our implementation, we set $M=10$ for the positional encoding. We employ a neural network with three fully connected layers for $\widetilde{g}_{\mathscr{E}, o}(\mathbf{\Theta}_{\mathscr{E}, o})$. Each hidden layer has 128 nodes and adopts the ReLU activation function. The neural object outputs a $48$-by-$1$ vector for the encoded representation of EM property $\bee$.

\subsubsection{Neural Interaction} The neural interaction models the interaction behavior, \textit{i.e.}, the transfer function $\widetilde{g}_{T}(\mathbf{\Theta}_T)$ in \eqref{eq:prob_form}. We propose to further divide the $\widetilde{g}_{T}(\mathbf{\Theta}_T)$ into individual neural networks, with each neural network modeling one type of interaction. In this work, we consider three common and important types of interactions: (i) reflection, (ii) diffraction, and (iii) diffuse scattering. Therefore, the $\widetilde{g}_{T}(\mathbf{\Theta}_T)$ incorporates three individual neural network models, which can be written as
\begin{align}
	\widetilde{g}_T(\mathbf{\Theta}_T) = \bigg\{\widetilde{g}_{T, \kappa}(\mathbf{\Theta}_{T, \kappa}) \mathrel{\Big|} \kappa=\{\mathrm{refl},\mathrm{diffr}, \mathrm{sact}\} \bigg\}.
\end{align}
We call the neural network model $\widetilde{g}_{T, \kappa}(\mathbf{\Theta}_{T, \kappa})$ that models the transfer function for an interaction type ``neural interaction".

In general, for a given type of interaction, the transfer function is determined by (i) the EM property of the interaction position $\bee$, (ii) the AoD and AoA of the incoming and outgoing EM waves $(\mathbf{d}^{\mathrm{AoA}}, \mathbf{d}^{\mathrm{AoD}})$, (iii) the interaction geometry $\cI_{l,i}$, and (iv) the incoming E-field $\bE_{l, i}$ in the case of nonlinear interactions. That is, \eqref{eq:prob_form} can be re-written as
\begin{align}\label{eq:26_}
	\bE_{l, i+1} = \widetilde{g}_{T, \kappa}\bigg(\bee, \mathbf{d}^{\mathrm{AoA}}_{l,i}, \mathbf{d}^{\mathrm{AoD}}_{l,i}, \cI_{l, i}, \bE_{l, i}; \mathbf{\Theta}_{T, \kappa}\bigg).
\end{align}
Note that, as the neural interaction model explicitly takes in the object's EM property $\bee$, the same neural interaction model can be shared across different objects.

\textbf{Polarization Direction Normalization.} An electric field $\bE$ can be decomposed as $\bE = E_{s} \hat{\bee}_s + E_p\hat{\bee}_p$, where $\hat{\bee}_s,\hat{\bee}_p \in \bbR^{3\times1}$ are unit vectors representing the two polarization directions, respectively. The $E_s, E_p\in \bbC$ are the complex amplitudes of the electric field along these polarization directions, respectively. Note that there are infinite ways of this orthogonal polarization decomposition, as long as $\hat{\bee}{s}$ and $\hat{\bee}{p}$ are orthogonal. Let $\bE = E_{s_1} \hat{\bee}_{s_1} + E_{p_1} \hat{\bee}_{p_1} = E_{s_2} \hat{\bee}{s_2} + E_{p_2} \hat{\bee}_{p_2}$ denote two arbitrary orthogonal decomposition of the polarization directions. Then, the complex amplitudes of the two decompositions are related as
\begin{equation}\label{eq:polar_convert}
	\left[\begin{array}{c}
		E_{s_2} \\
		E_{p_2}
	\end{array}\right]=\left[\begin{array}{cc}
		\hat{\mathbf{e}}_{\mathrm{i}, s_2}^T \hat{\mathbf{e}}_{\mathrm{i}, s_1} & \hat{\mathbf{e}}_{\mathrm{i}, s_2}^T \hat{\mathbf{e}}_{\mathrm{i}, p_1} \\
		\hat{\mathbf{e}}_{\mathrm{i}, \
			p_2}^T \hat{\mathbf{e}}_{\mathrm{i}, s_1} & \hat{\mathbf{e}}_{\mathrm{i}, p_2}^T \hat{\mathbf{e}}_{\mathrm{i}, p_1}
	\end{array}\right]\left[\begin{array}{c}
		E_{s_1} \\
		E_{p_1}
	\end{array}\right].
\end{equation}

To make the training of the neural interaction easier and more stable, we pre-define a unique orthogonal polarization decomposition for the incoming and outgoing E-field for a given interaction geometry and interaction type. For an incoming E-field, we always convert (normalize) it to the predefined incoming polarization decomposition. We also let the neural interaction model always output the outgoing E-field $\bE_{l,i+1}$ in the predefined outgoing polarization decomposition. With the predefined polarization directions, the neural interaction $\widetilde{g}_{T, \kappa}$ only needs to predict the complex amplitude of the outgoing E-field. Thus, \eqref{eq:26_} can be written as
\begin{align}\label{eq:27_}
	[E_{\mathrm{o}, s}, E_{\mathrm{o}, p}]^T = \widetilde{g}_{T, \kappa}\bigg(\bee, \mathbf{d}^{\mathrm{AoA}}, \mathbf{d}^{\mathrm{AoD}}, \cI, E_{\mathrm{i}, s}, E_{\mathrm{i}, p}; \mathbf{\Theta}_{T, \kappa}\bigg),
\end{align}
where we use $\bE_\mathrm{i}$ and $\bE_\mathrm{o}$ to denote the incoming and outgoing E-field, omitting the interaction index $i$ and path index $l$ for simplicity.
The predefined polarization decomposition varies with different interaction types and will be explained, along with the neural interaction models, in the next paragraphs.

\textbf{Reflection.} We normalize the incoming E-field to the transverse electric (TE) and transverse magnetic (TM) polarization components with respect to the reflection surface. With the incoming direction of the E-field $\bd^\mathrm{AoA}$ and the surface normal $\bn$ provided by $\cG$ in the geometric ray tracing, the TE and TM polarization directions can be written as
\begin{subequations}\label{eq:te_tm_pol}
	\begin{align}
		\hat{\mathbf{e}}_{\mathrm{i}, \perp} & =\frac{\bd^\mathrm{AoA} \times {\mathbf{n}}}{\left\|\bd^\mathrm{AoA} \times {\mathbf{n}}\right\|} \\
		\hat{\mathbf{e}}_{\mathrm{i}, \|} & =\hat{\mathbf{e}}_{i, \perp} \times \bd^\mathrm{AoA}.
	\end{align}
\end{subequations}
Similarly, we set the predefined polarization directions of the outgoing E-field as
\begin{subequations}
	\begin{align}
		& \hat{\mathbf{e}}_{\mathrm{o}, \perp}=\hat{\mathbf{e}}_{\mathrm{i}, \perp} \\
		& \hat{\mathbf{e}}_{\mathrm{o}, \|}=\frac{\hat{\mathbf{e}}_{\mathrm{r}, \perp} \times \bd^\mathrm{AoD}}{\left\|\hat{\mathbf{e}}_{\mathrm{r}, \perp} \times \bd^\mathrm{AoD}\right\|}.
	\end{align}
\end{subequations}

When the incoming EM wave hits an infinitely large plane of uniform non-magnetic material, it can be directly derived from Fresnel equation that the outgoing and incoming E-fields can be related by a scaling (diagonal) transfer matrix. Inspired by that, we design the neural interaction model for reflection $\widetilde{g}_{T, \mathrm{refl}}(\mathbf{\Theta}_{T, \mathrm{refl}})$ to predict a diagonal transfer matrix $\bT_\mathrm{refl}$. Let $\theta\in[0,\pi/2]$ denote the angle between the normal of the reflection surface $\bn$ and the incoming angle of the E-field $\bd^\mathrm{AoA}$. Considering the symmetricity of an infinitely large plane with uniform material, the transfer matrix only depends on the $\theta$ and the EM property $\bee$. Therefore, we design the neural interaction model for reflection to take in the angle $\theta$ and EM property $\bee$ and output a four-dimensional vector for the phase and amplitude of the diagonal components of $\bT_\mathrm{refl}$. In particular, we design the input to the neural interaction model as
\begin{align}\label{eq:refl_in}
	\bv_\mathrm{refl} = \big[\bee^T, \theta, \mathrm{PosEnc}(\bar{\theta})\big]^T,
\end{align}
where $\bar{\theta}$ is obtained by normalizing $\theta$ with the following function
\begin{align}\label{eq:angle_norm}
	f_\theta(x) = x / \pi - 1,
\end{align}
which normalizes an angle in $[0,2\pi]$ to [-1,1].

The output of the neural interaction model can be written as
\begin{align}\label{eq:refl_model}
	[t_1, t_2, t_3, t_4]^T = \widetilde{g}_{T, \mathrm{refl}}(\bv_\mathrm{refl};\mathbf{\Theta}_{T, \mathrm{refl}}).
\end{align}
The diagonal transfer matrix can then be obtained by
\begin{align}
	\bT_\mathrm{refl}=\left[\begin{array}{ll}
		t_1e^{jt_2} & 0 \\
		0 & t_3e^{jt_4}
	\end{array}\right].
\end{align}

In our implementation, we use $M=4$ for the positional encoding in \eqref{eq:refl_in}. The neural network $\widetilde{g}_{T, \mathrm{refl}}(\mathbf{\Theta}_{T, \mathrm{refl}})$ employs four fully connected layers with 128 nodes each. The hidden layers adopt the ReLU activation function. For the output layer, we adopt the activation functions $f_1(\cdot)$ and $f_2(\cdot)$ to the amplitudes ($t_1$ and $t_3$) and phases ($t_2$ and $t_4$), respectively:
\begin{align}
	f_1(x) &= e^x\\
	f_2(x) &= 2\pi \cdot \mathrm{sigmoid}(x).
\end{align}
With $f_1()$ and $f_2()$, the amplitude and phase are restricted to $[0, +\infty)$ and $[0, 2\pi]$, respectively.

\textbf{Diffraction.} We first normalize the incoming E-field with respect to the diffraction wedge. The predefined polarization directions of the incoming E-field are given by
\begin{subequations}
	\begin{align}
		\hat{\mathbf{e}}_{\mathrm{i}, s} & =\frac{\bd^\mathrm{AoA} \times \hat{\mathbf{e}}}{\left\|\bd^\mathrm{AoA} \times \hat{\mathbf{e}}\right\|} \\
		\hat{\mathbf{e}}_{\mathrm{i}, p} & =\hat{\mathbf{e}}_{\mathrm{i}, s} \times \bd^\mathrm{AoA},
	\end{align}
\end{subequations}
where $\hat{\bee}=\bn^0 \times \bn^n$ is the edge vector with $\bn^0, \bn^n$ representing the unit normal vectors of the two surfaces of the wedge. Similarly, the predefined polarization directions of the outgoing E-field are designed as
\begin{subequations}
	\begin{align}
		\hat{\mathbf{e}}_{\mathrm{o}, s} & = -\frac{\bd^\mathrm{AoD} \times \hat{\mathbf{e}}}{\left\|\bd^\mathrm{AoD} \times \hat{\mathbf{e}}\right\|} \\
		\hat{\mathbf{e}}_{\mathrm{o}, p} & =\hat{\mathbf{e}}_{\mathrm{r}, s} \times \bd^\mathrm{AoD}.
	\end{align}
\end{subequations}

In \cite{kouyoumjian1974uniform, holm2000new}, the authors proposed the UTD that models the diffraction effect at infinitely long wedges. The UTD shows that the outgoing and incoming E-fields can be related by a linear transfer matrix. To that end, we design the neural interaction model for diffraction $\widetilde{g}_{T,\mathrm{diffr}}(\mathbf{\Theta}_{T,\mathrm{diffr}})$ to predict a linear transfer matrix $\bT_{\mathrm{diffr}}$. Similar to the $\theta$ in reflection, we convert the incoming and outgoing directions $\bd^\mathrm{AoA}, \bd^\mathrm{AoD}$ to angle features $\beta_0, \beta_0^\prime, \phi, \phi^\prime$ relative to the wedge, which are given by
\begin{align}
	&\beta_0^\prime=\arccos(\hat{\bee}^T\bd^\mathrm{AoA})\label{eq:55}\\
	&\beta_0=\arccos(\hat{\bee}^T\bd^\mathrm{AoD})\label{eq:56}\\
	&\phi^{\prime} =\pi-\left[\pi-\cos ^{-1}\left((\bt_0)^T\bd_t^\mathrm{AoA} \right)\right] \operatorname{sgn}\left((\bt_0)^T\bd_t^\mathrm{AoA} \right) \label{eq:66} \\
	&\phi = \pi-\left[\pi-\cos ^{-1}\left((\bt_0)^T\bd_t^\mathrm{AoD} \right)\right] \operatorname{sgn}\left((\bt_0)^T\bd_t^\mathrm{AoD} \label{eq:67} \right),
\end{align}
where
\begin{align}
	\bt_0&=\bn^0 \times \hat{\mathbf{e}} \\
	\bd_t^\mathrm{AoA}&=\frac{\bd^\mathrm{AoA}-\left(\hat{\mathbf{e}}^T\bd^\mathrm{AoA} \right) \hat{\mathbf{e}}}{\left\|\bd^\mathrm{AoA}-\left(\hat{\mathbf{e}}^T \bd^\mathrm{AoA} \right) \hat{\mathbf{e}}\right\|} \\
	\bd_t^\mathrm{AoD}&=\frac{\bd^\mathrm{AoD}-\left(\hat{\mathbf{e}}^T\bd^\mathrm{AoD} \right) \hat{\mathbf{e}}}{\left\|\bd^\mathrm{AoD}-\left(\hat{\mathbf{e}}^T \bd^\mathrm{AoD} \right) \hat{\mathbf{e}}\right\|}.
\end{align}
The $\mathrm{sgn}(\cdot)$ is the sign function defined as
\begin{align}
	\operatorname{sgn}(x)= \begin{cases}1 , &x \geq 0 \\ -1 , &x<0\end{cases}
\end{align}
Moreover, we define $\beta_n$ as the exterior angle of the wedge, which is given by
\begin{align}
	\beta_n = 2\pi - \arccos\left(\bn_0^T\bn_n\right).
\end{align}

We design the input to the neural interaction model as
\begin{equation}
	\bv_\mathrm{diffr} = [\bee^T, \overline{\bv}_\mathrm{diffr}^T, \mathrm{PosEnc}^T(\overline{\bv}_\mathrm{diffr})]^T,\label{eq:v_diffr}
\end{equation}
where $\overline{\bv}_\mathrm{diffr}$ is given by
\begin{equation}
	\overline{\bv}_\mathrm{diffr} = [f_\theta(\phi^\prime), f_\theta(\phi), f_\theta(\beta_0)^\prime, f_\theta(\beta_0), f_\theta(\beta_n), \bar{s}^\prime, \bar{s}]^T. \label{eq:v_diffr__}
\end{equation}
The $\bar{s}^\prime, \bar{s}$ are the distances between the diffraction position to the last and next interaction position normalized by the maximum dimension of the scene. Note that $\overline{\bv}_\mathrm{diffr}$ is sufficient to uniquely define the geometry of the diffraction. Thus, combined with $\bee$, it can be used as the input to the neural interaction model for diffraction.

The output of the neural interaction model is an eight-dimensional vector corresponding to the amplitude and phases of the transfer matrix $\bT_\mathrm{diffr}$, which is given by
\begin{align}\label{eq:36}
	[t_1, t_2, t_3, t_4, t_5, t_6, t_7, t_8]^T &=\widetilde{g}_{T,\mathrm{diffr}}(\bv_\mathrm{diffr};\mathbf{\Theta}_{T,\mathrm{diffr}}).
\end{align}
The transfer matrix can then be obtained by
\begin{align}
	\bT_\mathrm{diffr}&=\left[\begin{array}{ll}
		t_1e^{jt_2} & t_3e^{jt_4} \\
		t_5e^{jt_6} & t_7e^{jt_8}
	\end{array}\right].
\end{align}

The neural network design of $\widetilde{g}_{T,\mathrm{diffr}}(\mathbf{\Theta}_{T,\mathrm{diffr}})$ employs five fully connected layers with 128 nodes each. The hidden layers adopt the ReLU activation function. For the output layer, we adopt the same activation functions to the amplitudes and phases as the reflection.

\textbf{Scattering.} We apply the same normalization to the polarization directions to the incoming E-field as reflection. We set the predefined polarization directions of the outgoing E-field $\hat{\mathbf{e}}_{\mathrm{o}, s}, \hat{\mathbf{e}}_{\mathrm{o}, p}$ as the spherical unit vectors of the outgoing direction $\bd^{AoD}$. Let $\bd^{AoD}=[x,y,z]^T$, $\hat{\mathbf{e}}_{\mathrm{o}, s}$ and $ \hat{\mathbf{e}}_{\mathrm{o}, p}$ are then given by
\begin{subequations}
	\begin{align}
		\hat{\mathbf{e}}_{\mathrm{o}, s} &= [\cos (\theta_s) \cos (\varphi_s), \cos (\theta_s) \sin (\varphi_s), -\sin (\theta_s)]^T\\
		\hat{\mathbf{e}}_{\mathrm{o}, p} &= [\sin (\varphi_s), \cos (\varphi_s), 0]^T,
	\end{align}
\end{subequations}
where $\theta_s=\arccos(z)$ and $\varphi_S=\operatorname{arctan}(y/x)$.

The current mathematical models for diffuse scattering are often developed with heuristic and empirical approaches. In \cite{degli2007measurement, degli2011analysis}, the authors propose a nonlinear model for the diffuse scattering effect and show some consistency between the proposed model and field measurement. In particular, the mathematical model in \cite{degli2007measurement} incorporates hypothesized scattering patterns. In the real world, the explicit expression of the scattering pattern is difficult to characterize and may also involve non-linearity with the E-field. To that end, we design the neural interaction model for scattering as a nonlinear transfer function.

Let $\mathbf{E}_{\mathrm{i}}=E_{\mathrm{i}, \perp} \hat{\mathbf{e}}_{\mathrm{i}, \perp}+E_{\mathrm{i}, \|} \hat{\mathbf{e}}_{\mathrm{i}, \|}$ denote the incoming E-field after the normalization. We design the input to the neural interaction model as \eqref{eq:v_scat_}.
\begin{figure*}
	\begin{equation}
		\bv_\mathrm{scat} = [\bee^T, \overline{\bv}_\mathrm{scat}^T, \mathrm{PosEnc}^T(\overline{\bv}_\mathrm{scat}), \Re\{\bar{E}_{\mathrm{i}, \perp}\}, \Im\{\bar{E}_{\mathrm{i}, \perp}\}, \Re\{\bar{E}_{\mathrm{i}, \|}\}, \Im\{\bar{E}_{\mathrm{i}, \|}\}]^T,\label{eq:v_scat_}
	\end{equation}
\end{figure*}

The $\overline{\bv}_\mathrm{scat}$ in \eqref{eq:v_scat_} is given by
\begin{equation}
	\overline{\bv}_\mathrm{scat} = \bigg[\big(\bd^\mathrm{AoA}\big)^T, \big(\bd^\mathrm{AoD}\big)^T, \bn \bigg]^T. \label{eq:v_scat__}
\end{equation}
The $\Re\{\cdot\}$ and $\Im\{\cdot\}$ in \eqref{eq:v_scat_} denote the real and imaginary parts of a complex scalar, respectively. The $\bar{E}_{\mathrm{i}, \perp}, \bar{E}_{\mathrm{i}, \|}$ are obtained by normalizing the incoming E-field following
\begin{align}
	\big[\bar{E}_{\mathrm{i}, \perp}, \bar{E}_{\mathrm{i}, \|}\big]^T = \frac{1}{\sqrt{|E_{\mathrm{i}, \perp}|^2+|E_{\mathrm{i}, \perp}|^2}} \big[E_{\mathrm{i}, \perp}, E_{\mathrm{i}, \|}\big]^T\label{eq:nonlinear_norm1}.
\end{align}

The output of the neural interaction model is a four-dimensional vector that directly predicts the amplitudes and phases of the outgoing E-field, which is given by
\begin{align}\label{eq:36}
	[t_1, t_2, t_3, t_4]^T &=\widetilde{g}_{T,\mathrm{scat}}(\bv_\mathrm{scat};\mathbf{\Theta}_{T,\mathrm{scat}}).
\end{align}
The predicted outgoing E-field is given by
\begin{align}
	\big[E_{\mathrm{s}, \theta}, E_{\mathrm{s}, \varphi}\big]^T = \sqrt{|E_{\mathrm{i}, \perp}|^2+|E_{\mathrm{i}, \perp}|^2} \, \big[t_1e^{jt_2}, t_3e^{jt_4}\big]^T\label{eq:scat_ml2}
\end{align}
\begin{figure*}[t]
	\centering
	\includegraphics[width=1.\linewidth]{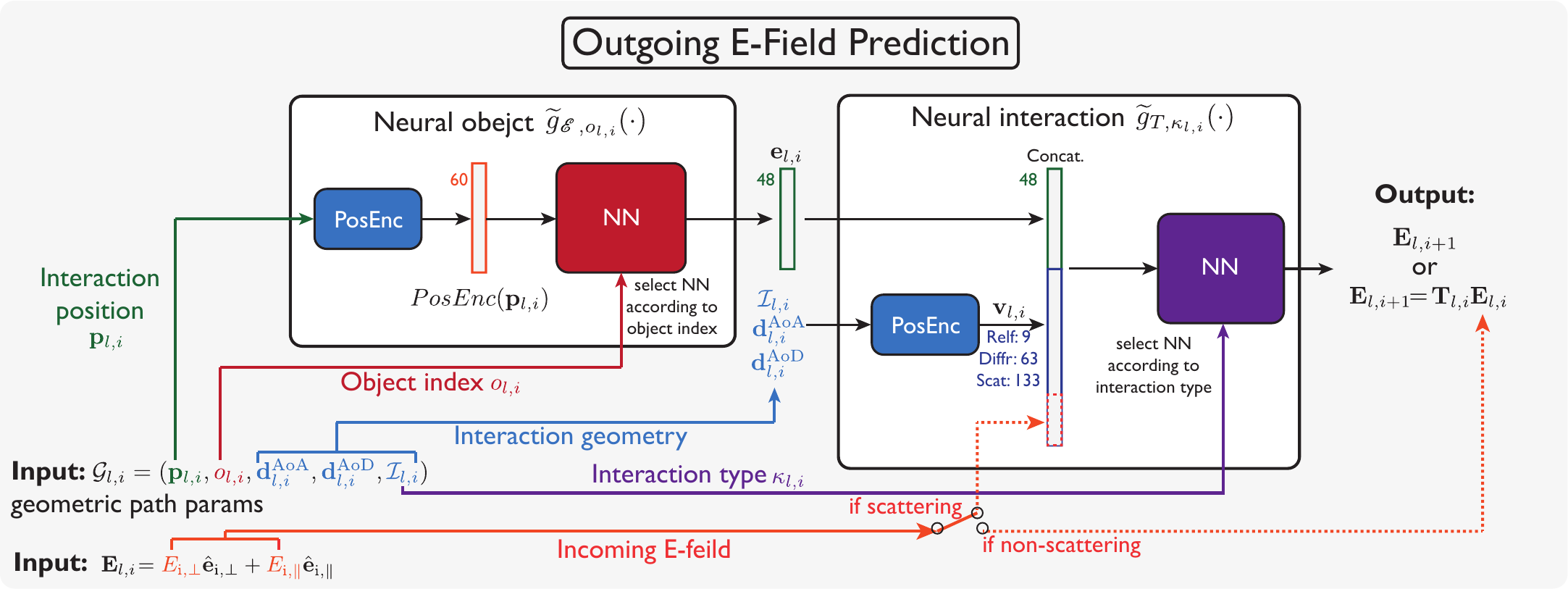}
	\caption{This figure illustrates the process of using the neural object and neural interaction models to predict the outgoing E-field of the $i$-th interaction in the $l$-th path. The E-field can be directly predicted by the transfer function (for scattering) or can be obtained by multiplying the incoming E-field by the predicted transfer matrix (for reflection and diffraction). The dimensions of the input, intermediate, and output vectors are annotated.}
	\label{fig:two_model}
\end{figure*}

The neural network design of $\widetilde{g}_{T,\mathrm{scat}}(\mathbf{\Theta}_{T,\mathrm{scat}})$ employs four fully connected layers with 128 nodes each. The hidden layers adopt the ReLU activation function. We adopt the same activation functions for the output layer to the amplitudes $t_1, t_3$ and phases $t_2, t_4$ as the reflection.

\begin{figure*}[t]
	\centering
	\includegraphics[width=1.\linewidth]{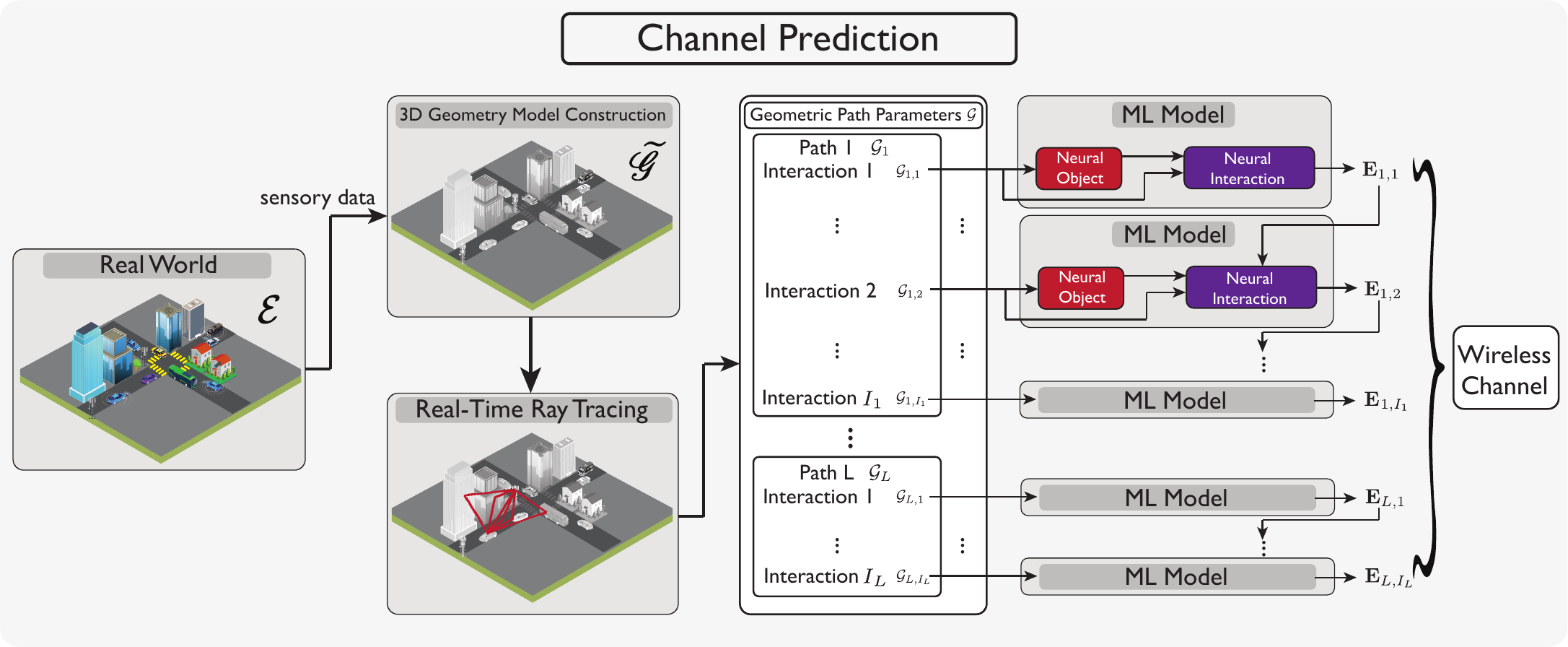}
	\caption{This figure shows the end-to-end process of using the learnable digital twin to predict the wireless channel. In particular, it first employs the neural object and neural interaction models to predict the outgoing E-fields of all paths, and then synthesize the channel with these E-fields.}
	\label{fig:end_to_end}
\end{figure*}

\section{End-to-End View}\label{sec:end_to_end}
This section provides an end-to-end view of the proposed learnable digital twin framework. In particular, we first show how the neural object and neural interaction models are combined to approximate the transfer function of an interaction and predict the outgoing E-field. After that, we elaborate on how the learnable digital can be used to predict the wireless channel. Lastly, we present the end-to-end training process of the learnable digital twin.
\subsection{Outgoing E-Field Prediction}\label{sec:Transfer Matrix Prediction}
\figref{fig:two_model} presents the process of using a neural object and neural interaction model to approximate the transfer function of an interaction. Let us first consider one interaction (\textit{e.g.} the $i$-th interaction in the $l$-th path) that is described by the geometric path parameter $\cG_{l, i} = (\bp_{l,i}, o_{l,i}, \mathbf{d}_{l, i}^{\mathrm{AoA}}, \mathbf{d}_{l, i}^{\mathrm{AoD}}, \cI_{l,i})$. As defined in \eqref{eq:geo_path_param}, $\mathbf{p}_{l, i}$ and $o_{l,i}$ denote the interaction position and the index of the interaction object, respectively. $\mathbf{d}_{l, i}^{\mathrm{AoA}}, \mathbf{d}_{l, i}^{\mathrm{AoD}}$ are unit vectors representing the incoming and outgoing ray directions. $\cI_{l,i}$ contains information about the interaction geometry and interaction type $\kappa_{l,i}\in\{\mathrm{refl}, \mathrm{diffr}, \mathrm{scat}\}$. We use the neural object and neural interaction models to approximate the transfer function and predict the outgoing E-field of this interaction. This process is summarized in the following steps.
\begin{itemize}
	\item \textbf{Step 1.} Select the neural object model according to the object index $o_{l,i}$.
	\item \textbf{Step 2.} Input the interaction position $\bp_{l,i}$ into the selected neural object model to obtain a high-dimensional representation of the EM property of this position.
	\item \textbf{Step 3.} Select the neural interaction model according to the interaction type $\kappa_{l,i}$ indicated in the interaction geometry $\cI_{l,i}$.
	\item \textbf{Step 4-1.} For linear interaction (reflection and diffraction), input the high-dimensional representation of the EM property $\bee$, AoA and AoD of the interaction $\mathbf{d}_{l, i}^{\mathrm{AoA}}, \mathbf{d}_{l, i}^{\mathrm{AoD}}$, and the interaction geometry $\cI_{l,i}$ to the selected interaction model. The interaction model outputs a transfer matrix $\bT_{l,i}$. The outgoing E-field can be obtained by $\bE_{l,i+1} = \bT_{l,i}\bE_{l,i}$.
	\item \textbf{Step 4-2.} For nonlinear interaction (diffuse scattering), input the high-dimensional representation of the EM property $\bee$, AoA and AoD of the interaction $\mathbf{d}_{l, i}^{\mathrm{AoA}}, \mathbf{d}_{l, i}^{\mathrm{AoD}}$, interaction geometry $\cI_{l,i}$, and the incoming E-field $\bE_{l,i}$ to the selected interaction model. The interaction model directly outputs the outgoing E-field $\bE_{l,i+1}$.
\end{itemize}

\subsection{Channel Prediction}\label{sec:Channel Prediction}
\figref{fig:two_model} presents the process of using a neural object and neural interaction models to predict the outgoing E-field of one interaction. Here, we show the procedure to predict the wireless channel using the proposed learnable digital twin framework, which is also illustrated in \figref{fig:end_to_end}. Let $\mathscr{G}$ denote the considered scenario. First, we construct (update) the geometry 3D model of the environment $\widetilde{\mathscr{G}}$, which incorporates $O$ objects. Then, we apply geometry ray tracing on the geometry 3D model $\widetilde{\mathscr{G}}$ to obtain the geometry path parameters of all paths and interactions $\cG$. After that, for each path, we iterate over the interactions along this path, and apply the corresponding neural object and neural interaction models to predict the outgoing E-field as discussed in \sref{sec:Transfer Matrix Prediction}. Lastly, the wireless channel can be synthesized with the predicted final outgoing E-field of all paths following \eqref{eq:cfr}, \eqref{eq:2}, and \eqref{eq:3}.

\subsection{End-to-End Training}
In \sref{sec:Transfer Matrix Prediction} and \sref{sec:Channel Prediction}, we elaborate on the channel prediction process assuming the neural objects and neural interactions are trained (\textit{i.e.} forward pass of the neural network). Here, we explain the end-to-end training process of the learnable digital twin.

\textbf{Data Collection.} Let $\mathscr{G}$ denote the considered real-world scenario. We collect two types of data from this real-world scenario: (i) the multi-modal sensing data and 6DoF pose information that are used to construct the geometry 3D model $\widetilde{\mathscr{G}}$, and (ii) the real-world communication channel $\bH$ coming from the crowd-sourced wireless samples. We apply the geometry ray tracing on the geometry 3D model $\widetilde{\mathscr{G}}$ to obtain the geometry path parameters $\cG$. With the geometry path parameters $\cG$ and the corresponding real-world channel, one data point (training sample) can be written as $(\cG, \bH)$. By changing the real-world scenario, \textit{e.g.}, changing the position of the receiver or the position/orientation of the other objects\footnote{Note that in real-world applications, the positions of the user receiver and other objects usually change by themselves with or without the need of training the digital twin. Crowd-sourced wireless samples are also often collected from different user positions.} We can collect a dataset $\cN = \{(\widetilde{\mathscr{G}}_n, \bH_n)\}_{n=1}^N$ with $N$ data points. Note that the object indices in $\widetilde{\mathscr{G}}_n$ need to be consistent across all data points.

\textbf{Training Process.} We initiate $O$ neural object models where $O$ is the number of all objects included in any $\widetilde{\mathscr{G}}_n \forall n\in N$. Furthermore, we initiate two neural interaction models for the reflection and diffraction. The (untrained) neural object models and neural interaction models are applied to predict the channel $\bH_n$ given $\widetilde{\mathscr{G}}_n$ as discussed in \sref{sec:Channel Prediction}. The neural object and interaction models can be trained by minimizing the following cost function.
\begin{align}
	\min_{
		\begin{subarray}{l}
			\{\widetilde{g}_{\mathscr{E}, o}(\mathbf{\Theta}_{\mathscr{E}, o}) | o=\{1,\hdots, O\} \}\\
			\{\widetilde{g}_{T, \kappa}(\mathbf{\Theta}_{T, \kappa}) | \kappa=\{\mathrm{refl},\mathrm{diffr}, \mathrm{sact}\}
		\end{subarray}
	} \quad	\sum_{n=1}^N\bigg[J\big(\bH_n, \widetilde{\bH}_n\big)\bigg].\label{eq:final_obj}
\end{align}
For the loss function $J$, we use
\begin{align}\label{eq:log_nmse_loss}
	J\big(\bH_n, \widetilde{\bH}_n\big) = \log_{10}\bigg(\frac{\|\bH_m-\widetilde{\bH}_n\|_F^2}{\|\bH_n\|_F^2}\bigg).
\end{align}
We used the TensorFlow 2.13 \cite{tensorflow} in our implementation to train the neural network models and used the Adam optimizer to minimize \eqref{eq:final_obj}.

\section{Experiment Setup}\label{sec:experiment_setup}
In this section, we first explain the considered scenario setup and then elaborate on the dataset generation process.
\subsection{Scenario Setup}
We consider two static scenarios: (i) a complex indoor scenario, \textit{Office}, to train the learnable digital twin and evaluate the channel prediction performance and (ii) a modified indoor scenario, \textit{Office-mod}, to demonstrate the evaluate the channel prediction performance under static environment changes. Next, we introduce the setup of these two scenarios.

\begin{figure}[t]
	\centering
	\includegraphics[width=1\linewidth]{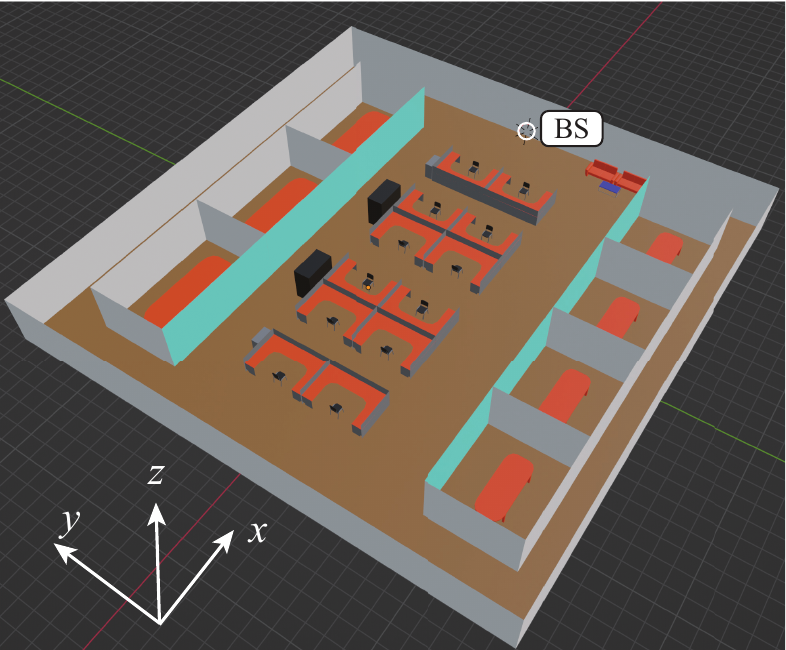}
	\caption{This figure shows the bird's view of the Office scenario.}
	\label{fig:office}
\end{figure}

\textbf{Office scenario.} We build the EM 3D model of the Office scenario using a 3D modeling software, Blender \cite{blender}. \figref{fig:office} shows the bird's view of the Office scenario, which reflects a complex indoor office room of size 30 meters by 30 meters. To demonstrate the proposed learnable digital twin's capability of learning the EM property and interaction behavior, we construct this complex Office scenario with multiple objects: walls, floor, desks, chairs, storages, conference tables, coffee tables, and sofas. The objects are made of seven different materials: concrete, wood, chipboard, plasterboard, metal, marble, and floorboard, each with a different EM property.

\begin{figure}[t]
	\centering
	\includegraphics[width=1\linewidth]{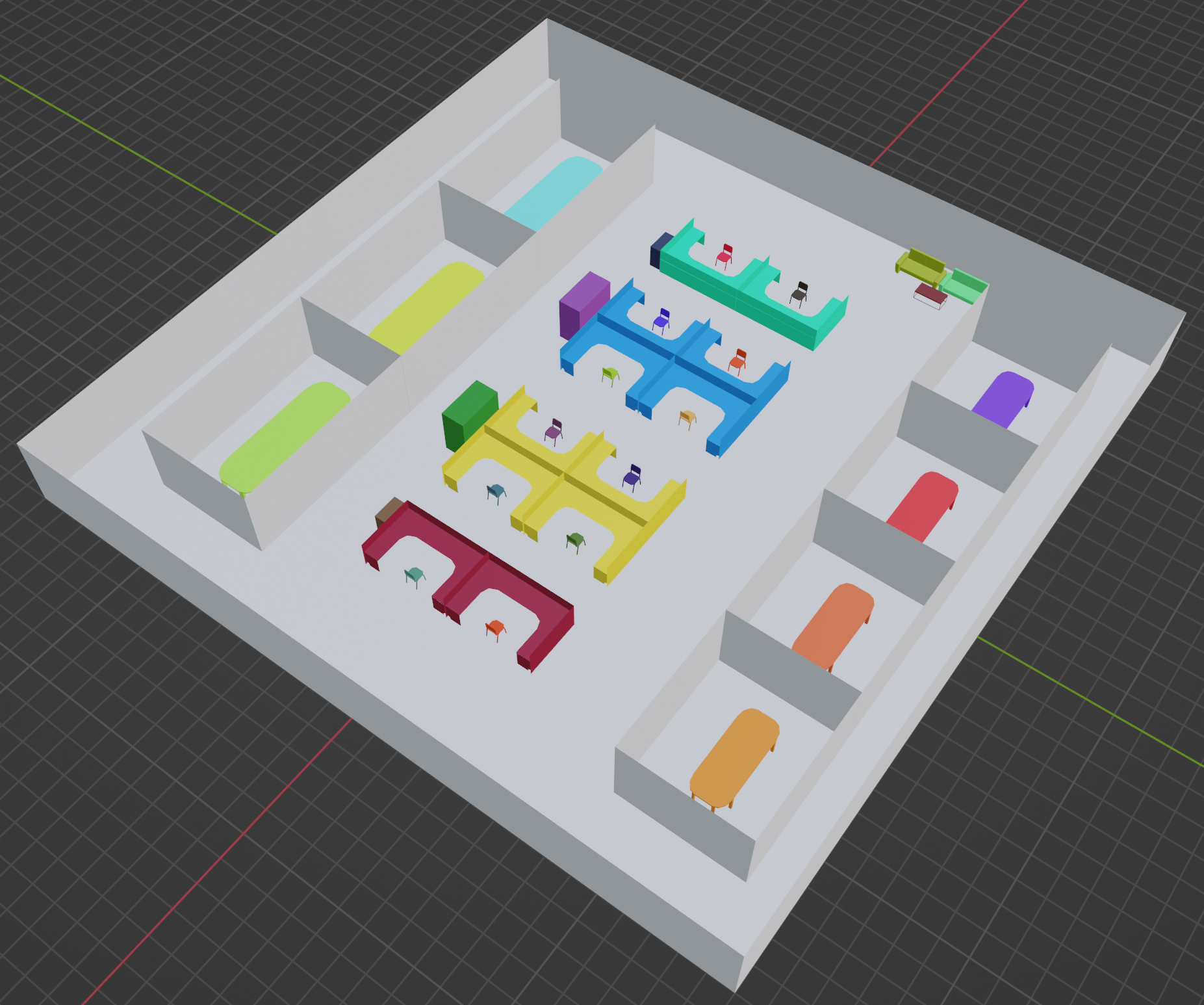}
	\caption{This figure annotates each object in the Office scenario in a different color. This figure aims to show the granularity of the 3D object detection to construct and update the digital twin. For instance, all the walls and the floor together form an object that is annotated in gray.}
	\label{fig:office_color_object}
\end{figure}
\begin{figure}[t]
	\centering
	\includegraphics[width=1\linewidth]{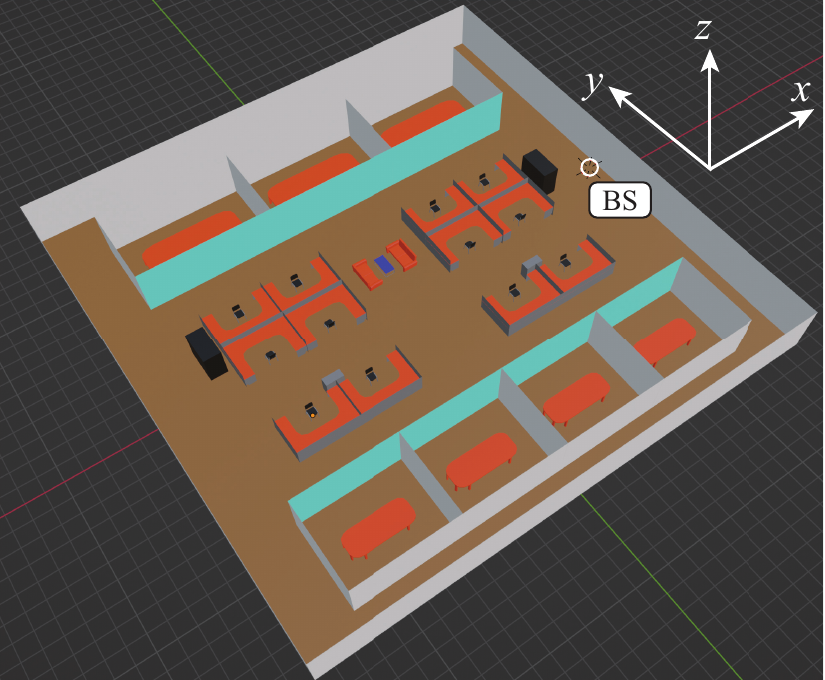}
	\caption{This figure shows the bird's view of the Office-mod scenario. This scenario consists of the same objects as the Office scenario but with a different layout.}
	\label{fig:office-mod}
\end{figure}

\textbf{Office-mod scenario.} In the real world, the geometry layout of the communication environment may change. For instance, an existing table can be moved to a different place. Therefore, it is essential for a trained digital twin to be able to predict the channel despite the changes in the environment. To that end, we construct the Office-mod scenario to incorporate environment changes. As shown in \figref{fig:office}, in the Office-mod scenario, we change the layout of the Office scenario in Blender by moving the positions of some of the existing objects except for the walls. Note that the walls can hardly be moved in real-world scenarios.

\begin{figure*}[h!]
	\centering
	\subfigure[Ground-truth room object]{\includegraphics[width = 0.45\linewidth]{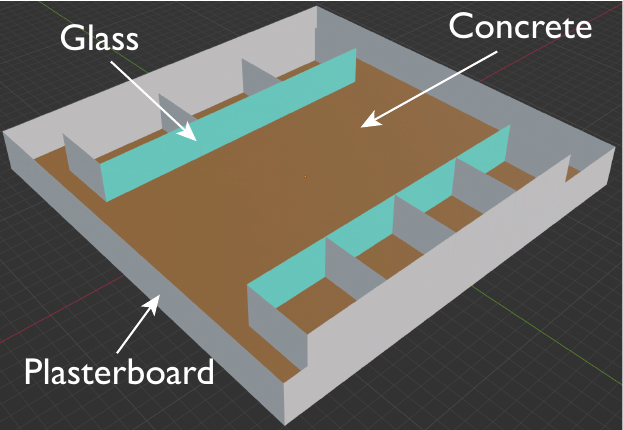}	\label{fig:room_mat_rep_a}}\quad
	\subfigure[Ground-truth table object]{\includegraphics[width = 0.45\linewidth]{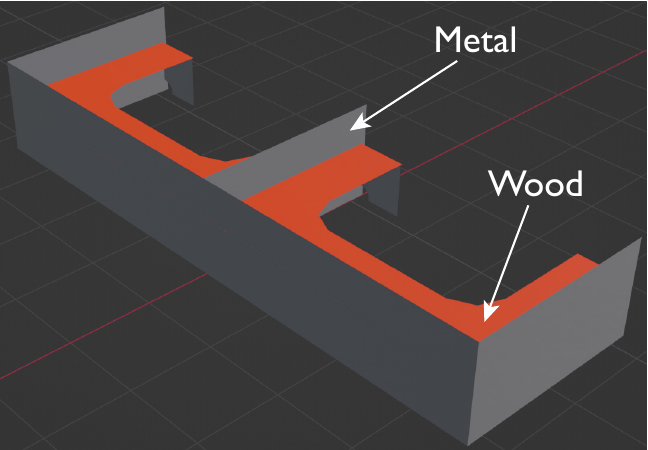}	\label{fig:table_mat_rep_b}}\\
	\subfigure[Learned EM representation of the room object (K-means clustering)]{\includegraphics[width = 0.44\linewidth]{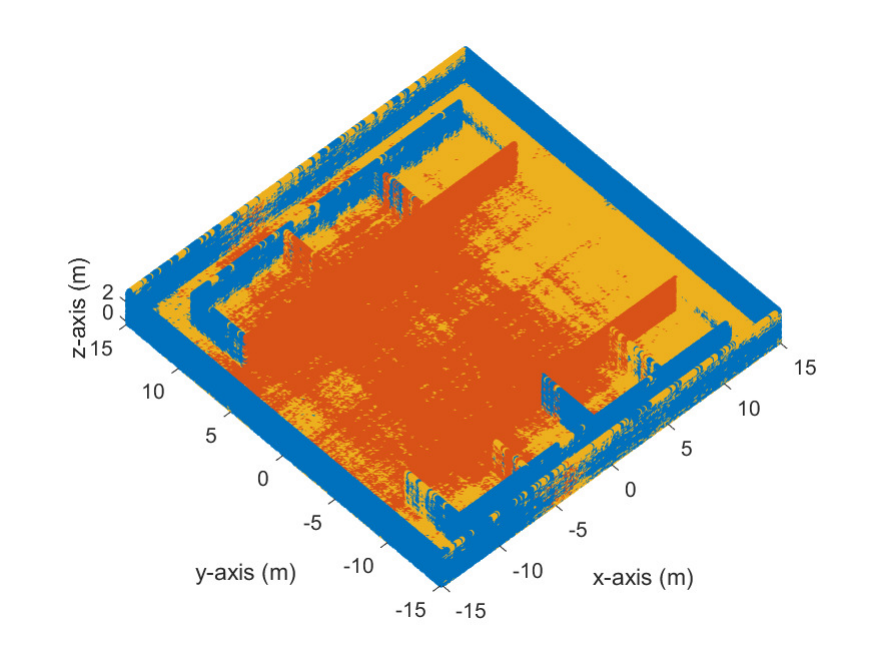} \label{fig:room_mat_rep_c}}\hfill
	\subfigure[Learned EM representation of the table object (K-means clustering)]{\includegraphics[width = 0.44\linewidth]{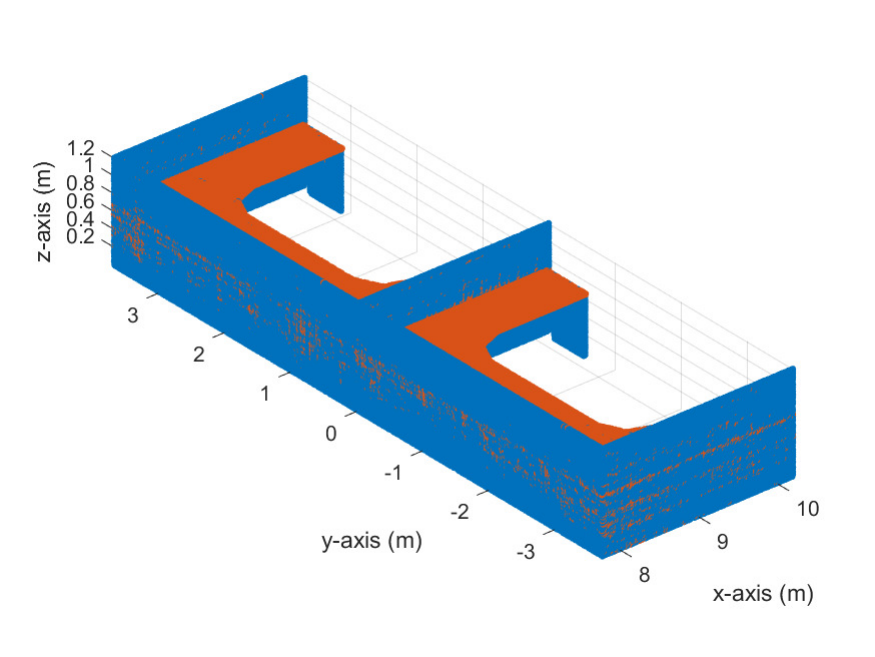} \label{fig:table_mat_rep_d}}\\
	\subfigure[Learned EM representation of the room object \mbox{(t-SNE)}]{\includegraphics[width = 0.44\linewidth]{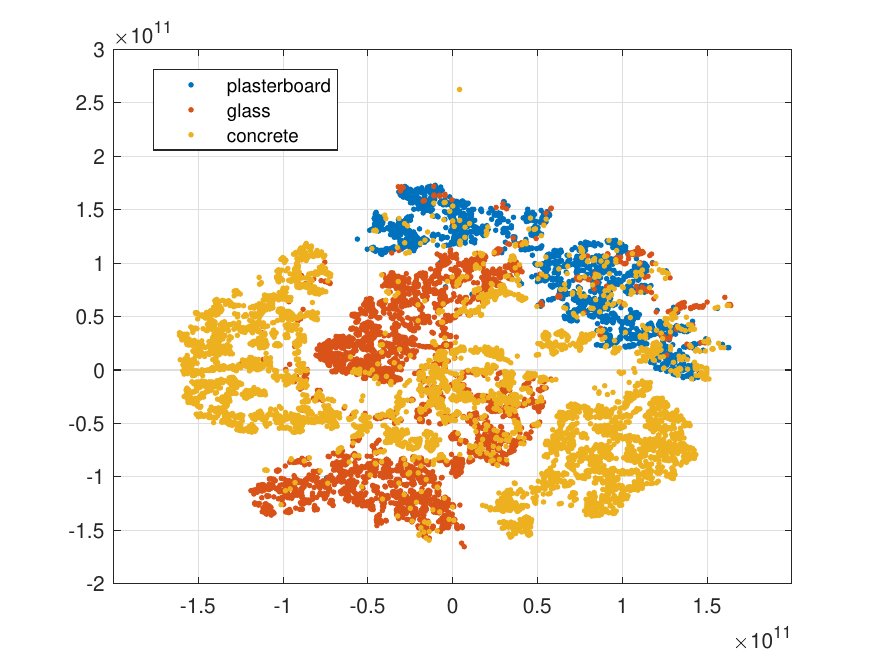} \label{fig:room_mat_rep_e}}\hfill
	\subfigure[Learned EM representation of the table object \mbox{(t-SNE)}]{\includegraphics[width = 0.44\linewidth]{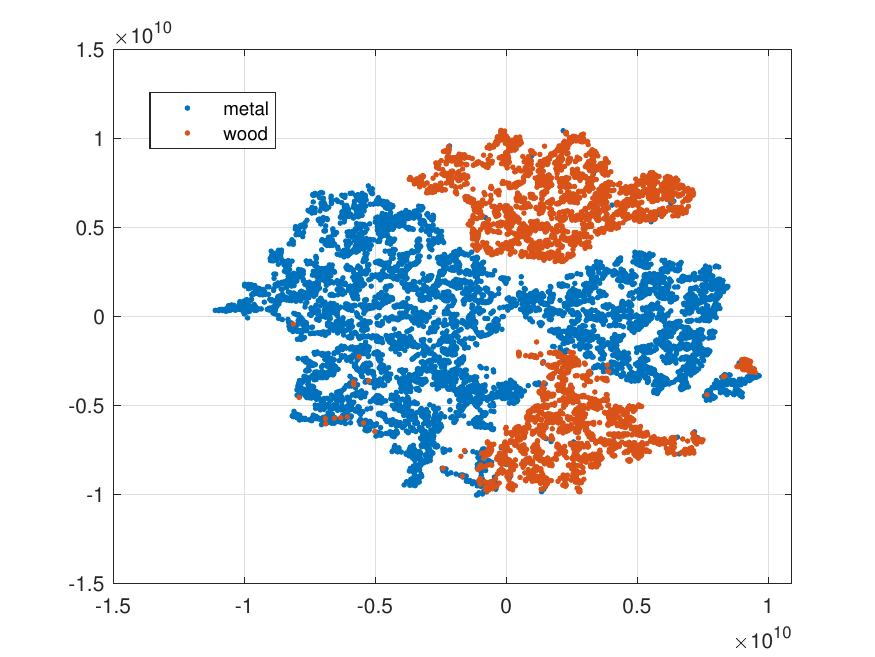} \label{fig:table_mat_rep_f}}
	\caption{This figure shows k-means clustering and t-SNE visualization of the learned high-dimensional EM property $\bee$ of the room and the table objects.}
	\label{fig:mat_rep}
\end{figure*}

\subsection{Data Generation}\label{sec:data generation}
This subsection explains the data generation process using the two scenarios.

\textbf{Office Scenario.} We apply Sionna 0.16 \cite{hoydis2022sionna} on the EM 3D model of the Office scenario to generate the datasets for training and testing the proposed learnable digital twin. Sionna is a TensorFlow-based open-source library for wireless ray tracing. In particular, given an EM 3D model, Sionna can trace propagation paths and provide the geometric propagation path parameters $\cG$ and the complex amplitude gain (\textit{i.e.} $\alpha_l$ in \eqref{eq:3}). In particular, Sionna tracks the following two types of paths: (i) a path consists of $I_\mathrm{refl}\geq 0$ reflections followed by $I_\mathrm{scat}\geq 0$ diffuse scattering, where $I_\mathrm{refl}+I_\mathrm{scat}\leq I_\mathrm{max}$ and (ii) a first-order diffraction path, \textit{i.e.}, the path goes through BS-diffraction-UE. We set the depth $I_\mathrm{max}=3$ in our implementation.

To generate the datasets, we first randomly sample $N$ UE receiver positions in the Office scenario. For each sample UE receiver position, we obtain the corresponding EM 3D model by including the UE receiver position in the geometry 3D map of the Office scenario. By excluding the material information, we obtain the geometric 3D model $\widetilde{\mathscr{G}}$ of each data point. After that, we apply Sionna ray tracing on the EM 3D model of each data point to obtain the geometric propagation path parameters $\cG$ and complex amplitude gains of all paths. Then, from the geometric propagation path parameters and complex amplitude gains, we generate the communication channel between the BS and the UE following \eqref{eq:2}. For simplicity, we consider single-antenna BS and single-antenna UE in the data generation process. However, it is worth noting that the proposed learnable digital twin can also be trained on or infer MIMO channels. We use 1024 subcarriers with $30$ kHz spacing for the OFDM signal. Therefore, the channel between the BS and UE can be written as $\bh\in\bbC^{1024\times 1}$. Let $\cG_m$ and $\bh_m$ denote the geometric propagation path parameters and communication channels for the $m$-th data point. The generated dataset can then be written as $\cM = \{(\cG_m, \bh_m)\}_{m=1}^M$. In our experiment, we generated $M=30000$ data points. We randomly select $25600$ and $1000$ data points to obtain the training and test datasets. The data points are selected so the training and test datasets do not overlap.

\textbf{Office-mod Scenario.} We apply the same process as the Office Scenario to generate a dataset using the EM 3D model of the Office-mod scenario. The dataset generated from the Office-mod scenario consists of $1000$ data points, which are used only for testing the learnable digital twin (\textit{i.e.} not for training).

\section{Simulation Results}\label{sec:result}
In this section, we aim to evaluate the performance of the learnable digital twin. First, we visualize the learned high-dimensional EM property. After that, we present the accuracy of the channel prediction.
\subsection{Does the Neural Object Learn the EM Property of Different Materials?}

For visualization, we randomly sample some positions on the surface of the objects in the Office scenario, and input the sampled positions into the corresponding neural object model to retrieve the learned high-dimensional EM property $\bee$. After that, we employ k-means clustering to cluster the high-dimensional EM property $\bee$ from each object. In \figref{fig:mat_rep}, we present the clustering results of the room and the table object and compare that with the ground-truth EM 3D models in the Office scenario. It can be seen that the positions sharing the same materials are grouped into the same cluster (\textit{i.e.} shown in the same color). \textbf{This indicates that the neural object models implicitly learn meaningful high-dimensional representations of the EM property and can distinguish different materials.}

For further visualization, we apply the t-SNE \cite{van2008visualizing} to the learned high-dimensional EM property $\bee$, and map it into the two-dimensional space. The t-SNE algorithm preserves the local structures and clusters similar data points in the high-dimensional space closer to the low-dimensional space. As depicted in \figref{fig:table_mat_rep_f}, the learned EM property $\bee$ corresponding to metal and wood materials can be well separated for the table object. However, \figref{fig:room_mat_rep_e} shows that, for the room object, points corresponding to the glass and concrete material are partially overlapping. It can also be observed from \figref{fig:room_mat_rep_c} that the positions of the concrete and glass materials are clustered into the same group near the junction of the glass wall and the concrete floor. Although the position is the input to the neural object model, the neural object model should ideally learn to output only the EM property and de-correlate the position information.

\subsection{Does the Digital Twin Accurately Predict the Channels?}
\begin{figure}[t]
	\centering
	\includegraphics[width=1\linewidth]{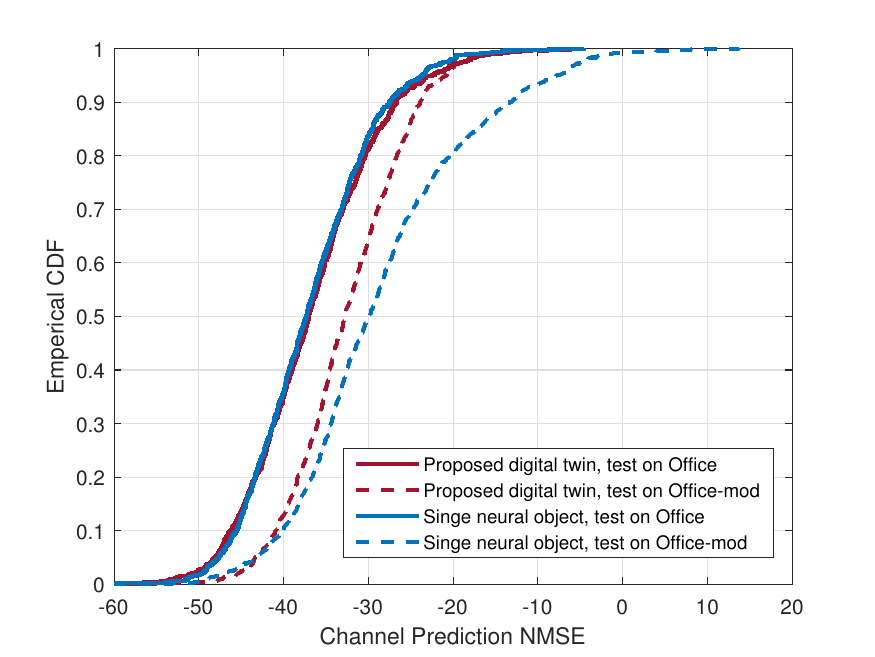}
	\caption{This figure shows the CDF of the channel prediction NMSE of learnable digital twin models trained on the Office scenario. The proposed learnable digital twin achieves high channel prediction performance on both the Office scenario and the Office-mod scenario with environment changes.}
	\label{fig:3}
\end{figure}

\begin{figure*}[h!]
	\centering
	\subfigure[Channel amplitude]{\includegraphics[width = 1.\linewidth]{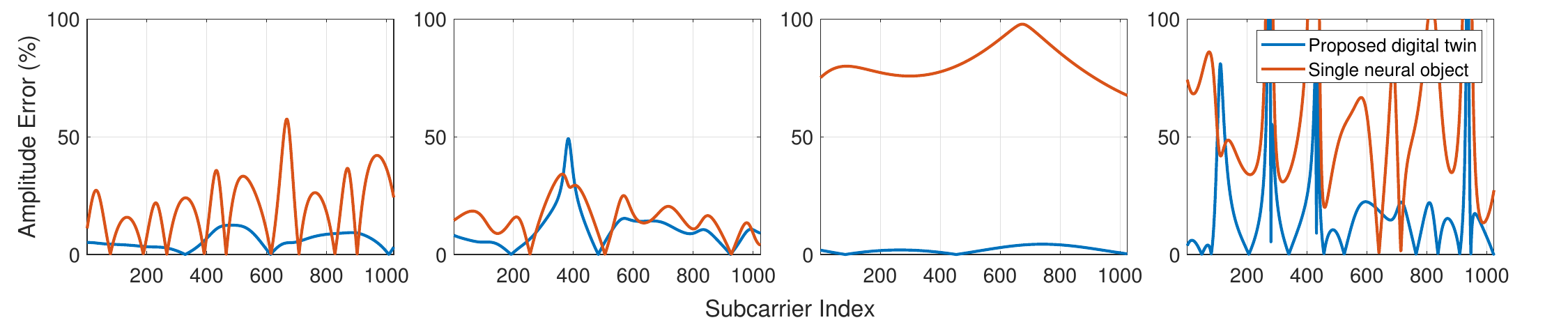}	\label{fig:office_mod1_channel_gain_comp}}\\
	\subfigure[Channel phase]{\includegraphics[width = 1.\linewidth]{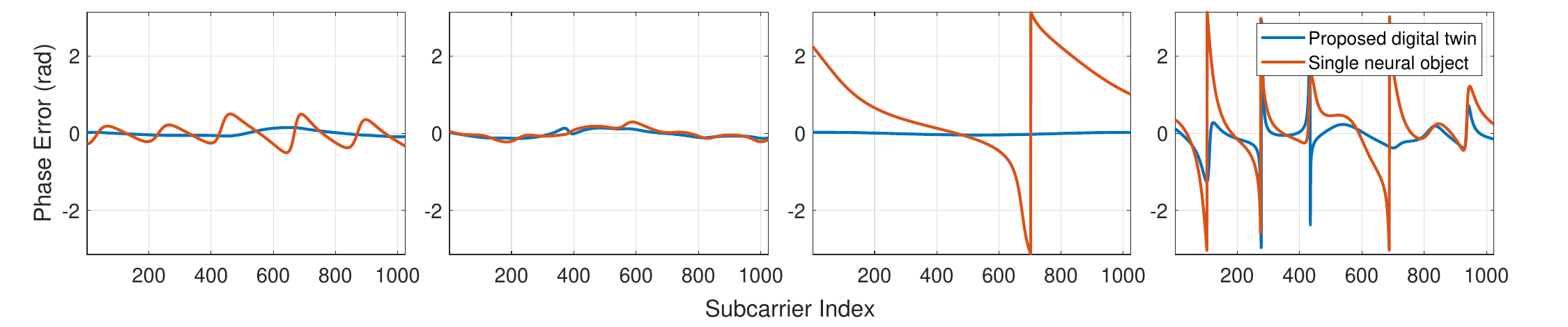}	\label{fig:office_mod1_channel_phase_comp}}
	\caption{This figure compares the channel gain and phases of the ground-truth channel and the predicted channel in Office-mod. The approaches using single or multiple neural objects are trained on the data from the Office scenario.}
	\label{fig:office_mod1_channel_all_comp}
\end{figure*}

\begin{figure}[h!]
	\centering
	\subfigure[Proposed digital twin (multiple neural objects)]{\includegraphics[width = 1\linewidth]{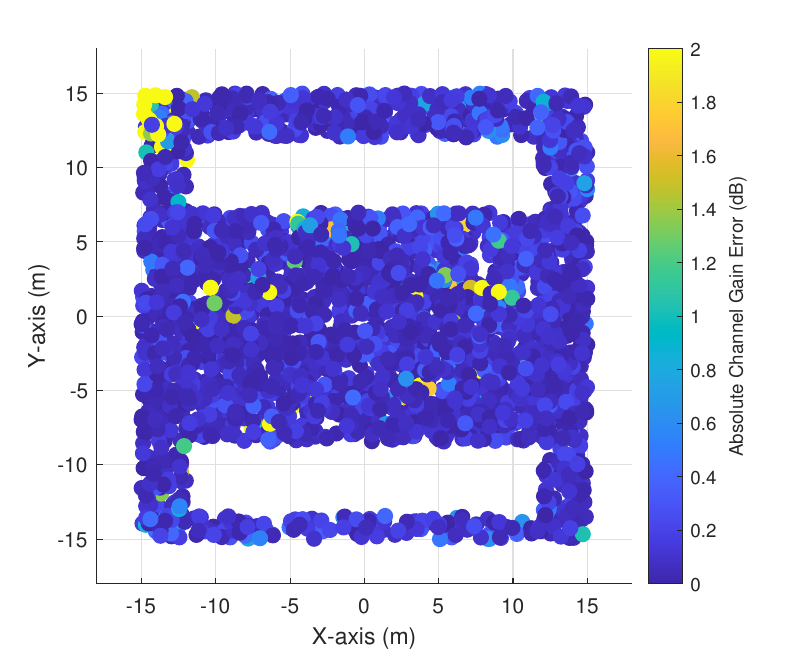}	\label{fig:office_mod1_channel_gain_vis_proposed}}
	\subfigure[Single neural object]{\includegraphics[width = 1\linewidth]{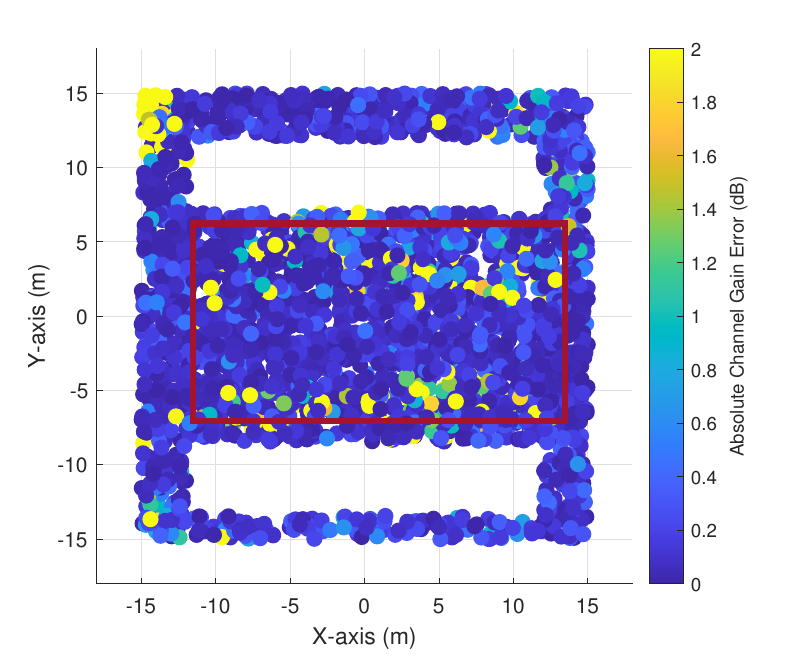}	\label{fig:office_mod1_channel_gain_vis_single}}
	\caption{This figure shows the absolute channel gain error by using single or multiple neural objects to model the EM property. The two approaches are trained on the data from the Office scenario and tested on the data from the Office-mod scenario.}
	\label{fig:office_channel_gain_vis_all}
\end{figure}
One key function of the digital twin is to predict the communication channel with reduced or even eliminated pilot training overhead. The accuracy of the predicted channel directly impacts the downstream communication tasks and determines the performance of the communication system. Therefore, in this subsection, we evaluate the accuracy of channel prediction. We adopt the channel NMSE as the performance metric. The channel NMSE evaluates how close an estimated/predicted channel is to the ground-truth channel. Let $\widetilde{\bH}$ denote the estimated/predicted channel tensor with arbitrary dimensions. Let $\bH$ denote the ground-truth channel. The channel NMSE can be written as
\begin{align}
	NMSE = \frac{\|flatten(\bH-\widetilde{\bH})\|^2_2}{\|flatten(\bH)\|^2_2},
\end{align}
where $flatten(\cdot)$ flattens the channel tensor into a one-dimensional vector, and $\|\cdot\|_2$ is the $l^2$ vector norm.

We train the learnable digital twin on the training dataset generated from the Office scenario, and test the digital twin on the test datasets generated from the Office scenario and the Office-mod scenario. \figref{fig:3} shows the channel prediction NMSE of the trained digital twin (labeled as ``proposed digital twin"). It can be seen that the trained digital twin can achieve low channel NMSE on the Office scenario (represented by the solid dark red color in the figure). \textbf{For example, for $\mathbf{90\%}$ of the UEs, the channel prediction NMSE is lower than $-26$ dB. Note that, once trained, the learnable digital twin can achieve this high channel prediction without any pilot training.}

\subsection{Does the Digital Twin Adapt to Environment Changes?}
The capability to adapt to environment changes is vital for the digital twins: Real-world communication environments often incorporate objects that can change their positions. Via crowd-sourced wireless samples, the proposed learnable digital twin is capable of learning the EM property of each object in the scene, and via ray tracing, the proposed method is capable of embedding the dynamics (\textit{i.e.}, the object position changes) into the end-to-end deep learning framework. To evaluate the capability of the learnable digital twin to handle environment changes, we train the learnable digital twin on the Office scenario and test it on the Office-mod scenario. As a baseline approach, we also evaluate a digital twin model that models the entire environment with one single neural object model\footnote{The neural object model in the baseline digital twin is adjusted such that the number of trainable parameters is similar to the total number of trainable parameters of all the neural object models in the proposed digital twin.}. This baseline digital twin model is also trained on the Office scenario and tested on the Office-mod scenario.

As shown in \figref{fig:3}, the proposed digital twin model can generalize to the Office-mod scenario and achieve only slightly higher channel prediction NMSE than that in the Office scenario. In particular, for $90\%$ of the UEs, the channel prediction NMSE is lower than $-20$ dB. Compared with the baseline digital twin (noted as ``single neural object"), the proposed digital twin achieves similar channel prediction performance in the Office scenario (without environment changes) as expected and \textbf{higher channel prediction performance} in the Office-mod scenario (with changes). In particular, we observe much more samples with large performance degradation (\textit{i.e.}, right-side tail of the dotted blue curve). In \figref{fig:office_mod1_channel_all_comp}, we show some randomly sampled channels from the Office-mod scenario and compare the amplitude and phase error of the predicted channels of the two digital twin approaches.
Let $\alpha, \widetilde{\alpha}>0$ denote the ground-truth and the predicted amplitudes of a subcarrier, respectively. The amplitude error on this subcarrier is then given by $|\widetilde{\alpha} - \alpha|/\alpha$. The phase error is the absolute phase difference between the ground-truth and the predicted channels wrapped to $[-\pi, \pi)$.
It can be seen that the proposed digital twin model with multiple neural objects better approximates the ground-truth channels. \textbf{The above results demonstrate that, when the objects in the scenario move, the proposed digital twin can efficiently adapt to the environment changes and generalize well without additional training/refining data.} This capability is essential for the digital twin to operate and aid communications in real-world scenarios where environmental changes are ubiquitous.

In \figref{fig:office_channel_gain_vis_all}, we plot the heatmaps of the absolute channel gain error of the proposed and the baseline digital twin models. The channel gain is calculated by summing the gain of all paths (or, equivalently, all subcarriers). Let $P_\mathrm{dB}$ and $\widetilde{P}_\mathrm{dB}$ denote the channel gains of the ground-truth channel and the predicted channels in dB values. The absolute channel gain error is given by $|\widetilde{P}_\mathrm{dB}-P_\mathrm{dB}|$. It can be seen from \figref{fig:office_channel_gain_vis_all} that, when tested on the Office-mod scenario, the baseline digital twin model leads to significantly larger error. In particular, the large-error positions are mostly located in the middle area of the office room (annotated in red). Comparing \fig{fig:office} and \fig{fig:office-mod}, it is evident that most of the environmental changes in the Office-mod scenario stem from the desks, chairs and cabinets in this area, partially explaining the performance degradation of the baseline digital twin model. Therefore, we anticipate that the proposed digital twin model will be more advantageous in more complex scenarios with more environment changes.

\section{Conclusion}\label{sec:Conclusion}
In this paper, we investigate a novel direction of digital twin-aided communications. Using the 3D geometry and EM model of the surrounding objects and ray tracing, the digital twin can predict the communication channels with reduced or eliminated pilot training overhead. The predicted channels can be used to solve various downstream communication tasks and achieve high system performance.

The key challenges of digital twin-aided communications lie in obtaining the EM property and modeling the interaction behavior of the communication environment. To that end, we propose a learnable digital twin framework that can learn the EM property and interaction behavior of the surrounding objects from the communication channels in an end-to-end manner. Moreover, the proposed learnable digital twin is designed to generalize to the environment changes in the scenario, which is essential when operating in most real-world scenarios.

Simulation results show that the proposed learnable digital twin can implicitly learn the EM property of the objects and distinguish between different materials. Moreover, the proposed learnable digital twin can predict the communication channels accurately, where the channel prediction NMSE is lower than $-26$ dB in $90\%$ cases. It is also demonstrated that the proposed learnable digital twin can generalize well to environment changes with a relatively small performance drop in channel prediction accuracy. Once trained, the learnable digital twin does not require any pilot training overhead to predict the channels. These results highlight the promising prospect of this novel direction.

\balance
\bibliographystyle{IEEEtran}

\begin{thebibliography}{10}
	\providecommand{\url}[1]{#1}
	\csname url@samestyle\endcsname
	\providecommand{\newblock}{\relax}
	\providecommand{\bibinfo}[2]{#2}
	\providecommand{\BIBentrySTDinterwordspacing}{\spaceskip=0pt\relax}
	\providecommand{\BIBentryALTinterwordstretchfactor}{4}
	\providecommand{\BIBentryALTinterwordspacing}{\spaceskip=\fontdimen2\font plus
		\BIBentryALTinterwordstretchfactor\fontdimen3\font minus
		\fontdimen4\font\relax}
	\providecommand{\BIBforeignlanguage}[2]{{%
			\expandafter\ifx\csname l@#1\endcsname\relax
			\typeout{** WARNING: IEEEtran.bst: No hyphenation pattern has been}%
			\typeout{** loaded for the language `#1'. Using the pattern for}%
			\typeout{** the default language instead.}%
			\else
			\language=\csname l@#1\endcsname
			\fi
			#2}}
	\providecommand{\BIBdecl}{\relax}
	\BIBdecl
	
	\bibitem{alkhateeb2023real}
	A.~Alkhateeb, S.~Jiang, and G.~Charan, ``{Real-time digital twins: Vision and
		research directions for 6G and beyond},'' \emph{IEEE Commun. Mag.}, 2023.
	
	\bibitem{boccardi2014five}
	F.~Boccardi, R.~W. Heath, A.~Lozano, T.~L. Marzetta, and P.~Popovski, ``Five
	disruptive technology directions for 5g,'' \emph{IEEE communications
		magazine}, vol.~52, no.~2, pp. 74--80, 2014.
	
	\bibitem{jiang2023digital}
	S.~Jiang and A.~Alkhateeb, ``Digital twin based beam prediction: Can we train
	in the digital world and deploy in reality?'' in \emph{2023 IEEE
		International Conference on Communications Workshops (ICC Workshops)}, 2023,
	pp. 36--41.
	
	\bibitem{jiang2024digitalcsi}
	------, ``{Digital Twin Aided Massive MIMO: CSI Compression and Feedback},'' in
	\emph{IEEE ICC}.\hskip 1em plus 0.5em minus 0.4em\relax IEEE, 2024.
	
	\bibitem{alrabeiah2019deep}
	M.~Alrabeiah and A.~Alkhateeb, ``Deep learning for tdd and fdd massive mimo:
	Mapping channels in space and frequency,'' in \emph{2019 53rd asilomar
		conference on signals, systems, and computers}.\hskip 1em plus 0.5em minus
	0.4em\relax IEEE, 2019, pp. 1465--1470.
	
	\bibitem{hua2018accurate}
	J.~Hua, H.~Sun, Z.~Shen, Z.~Qian, and S.~Zhong, ``Accurate and efficient
	wireless device fingerprinting using channel state information,'' in
	\emph{IEEE INFOCOM 2018-IEEE Conference on Computer Communications}.\hskip
	1em plus 0.5em minus 0.4em\relax IEEE, 2018, pp. 1700--1708.
	
	\bibitem{wu2012csi}
	K.~Wu, J.~Xiao, Y.~Yi, D.~Chen, X.~Luo, and L.~M. Ni, ``Csi-based indoor
	localization,'' \emph{IEEE Transactions on Parallel and Distributed Systems},
	vol.~24, no.~7, pp. 1300--1309, 2012.
	
	\bibitem{wang2016csi}
	X.~Wang, L.~Gao, and S.~Mao, ``Csi phase fingerprinting for indoor localization
	with a deep learning approach,'' \emph{IEEE Internet of Things Journal},
	vol.~3, no.~6, pp. 1113--1123, 2016.
	
	\bibitem{jagannath2022comprehensive}
	A.~Jagannath, J.~Jagannath, and P.~S. P.~V. Kumar, ``A comprehensive survey on
	radio frequency (rf) fingerprinting: Traditional approaches, deep learning,
	and open challenges,'' \emph{Computer Networks}, vol. 219, p. 109455, 2022.
	
	\bibitem{koike2020fingerprinting}
	T.~Koike-Akino, P.~Wang, M.~Pajovic, H.~Sun, and P.~V. Orlik,
	``Fingerprinting-based indoor localization with commercial mmwave wifi: A
	deep learning approach,'' \emph{IEEE Access}, vol.~8, pp. 84\,879--84\,892,
	2020.
	
	\bibitem{yun2015ray}
	Z.~Yun and M.~F. Iskander, ``Ray tracing for radio propagation modeling:
	Principles and applications,'' \emph{IEEE access}, vol.~3, pp. 1089--1100,
	2015.
	
	\bibitem{geok2018comprehensive}
	T.~K. Geok, F.~Hossain, M.~Kamaruddin, N.~Rahman, S.~Thiagarajah, A.~T.~W.
	Chiat, and C.~Liew, ``A comprehensive review of efficient ray-tracing
	techniques for wireless communication,'' \emph{International Journal on
		Communications Antenna and Propagation}, vol.~8, no.~2, pp. 123--136, 2018.
	
	\bibitem{deng2017toward}
	Y.~Deng, Y.~Ni, Z.~Li, S.~Mu, and W.~Zhang, ``Toward real-time ray tracing: A
	survey on hardware acceleration and microarchitecture techniques,'' \emph{ACM
		Computing Surveys (CSUR)}, vol.~50, no.~4, pp. 1--41, 2017.
	
	\bibitem{meister2021survey}
	D.~Meister, S.~Ogaki, C.~Benthin, M.~J. Doyle, M.~Guthe, and J.~Bittner, ``A
	survey on bounding volume hierarchies for ray tracing,'' in \emph{Computer
		Graphics Forum}, vol.~40, no.~2.\hskip 1em plus 0.5em minus 0.4em\relax Wiley
	Online Library, 2021, pp. 683--712.
	
	\bibitem{kouyoumjian1974uniform}
	R.~G. Kouyoumjian and P.~H. Pathak, ``A uniform geometrical theory of
	diffraction for an edge in a perfectly conducting surface,''
	\emph{Proceedings of the IEEE}, vol.~62, no.~11, pp. 1448--1461, 1974.
	
	\bibitem{holm2000new}
	P.~D. Holm, ``A new heuristic utd diffraction coefficient for non-perfectly
	conducting wedges,'' \emph{IEEE Transactions on Antennas and Propagation},
	vol.~48, no.~8, pp. 1211--1219, 2000.
	
	\bibitem{degli2007measurement}
	V.~Degli-Esposti, F.~Fuschini, E.~M. Vitucci, and G.~Falciasecca, ``Measurement
	and modelling of scattering from buildings,'' \emph{IEEE transactions on
		antennas and propagation}, vol.~55, no.~1, pp. 143--153, 2007.
	
	\bibitem{orekondy2023winert}
	T.~Orekondy, P.~Kumar, S.~Kadambi, H.~Ye, J.~Soriaga, and A.~Behboodi, ``WiNeRT: Towards neural ray tracing for wireless channel modelling and differentiable simulations,'' \emph{The Eleventh International Conference On Learning Representations}, 2023. 
	
	\bibitem{hoydis2023learning}
	J.~Hoydis, F.~Aoudia, S.~Cammerer, F.~Euchner, M.~Nimier-David, S.~Brink, and A.~Keller, ``Learning radio environments by differentiable ray tracing,'' \emph{arXiv Preprint, arXiv:2311.18558}, 2023.
	
	\bibitem{posenc}
	B.~Mildenhall, P.~Srinivasan, M.~Tancik, J.~Barron, R.~Ramamoorthi, and R.~N.
	Nerf, ``Representing scenes as neural radiance fields for view synthesis.,
	2021, 65,'' \emph{DOI: https://doi. org/10.1145/3503250}, pp. 99--106.
	
	\bibitem{muller2022instant}
	T.~M{\"u}ller, A.~Evans, C.~Schied, and A.~Keller, ``Instant neural graphics
	primitives with a multiresolution hash encoding,'' \emph{ACM transactions on
		graphics (TOG)}, vol.~41, no.~4, pp. 1--15, 2022.
	
	\bibitem{zhao2023nerf2}
	X.~Zhao, Z.~An, Q.~Pan, and L.~Yang, ``Nerf2: Neural radio-frequency radiance
	fields,'' in \emph{Proceedings of the 29th Annual International Conference on
		Mobile Computing and Networking}, 2023, pp. 1--15.
	
	\bibitem{nie2020total3dunderstanding}
	Y.~Nie, X.~Han, S.~Guo, Y.~Zheng, J.~Chang, and J.~J. Zhang,
	``Total3dunderstanding: Joint layout, object pose and mesh reconstruction for
	indoor scenes from a single image,'' in \emph{Proceedings of the IEEE/CVF
		Conference on Computer Vision and Pattern Recognition}, 2020, pp. 55--64.
	
	\bibitem{zhang2021holistic}
	C.~Zhang, Z.~Cui, Y.~Zhang, B.~Zeng, M.~Pollefeys, and S.~Liu, ``Holistic 3d
	scene understanding from a single image with implicit representation,'' in
	\emph{Proceedings of the IEEE/CVF Conference on Computer Vision and Pattern
		Recognition}, 2021, pp. 8833--8842.
	
	\bibitem{zhang2021deeppanocontext}
	C.~Zhang, Z.~Cui, C.~Chen, S.~Liu, B.~Zeng, H.~Bao, and Y.~Zhang,
	``Deeppanocontext: Panoramic 3d scene understanding with holistic scene
	context graph and relation-based optimization,'' in \emph{Proceedings of the
		IEEE/CVF International Conference on Computer Vision}, 2021, pp.
	12\,632--12\,641.
	
	\bibitem{qi_link3}
	``{SceneScript: an AI model and method to understand and describe 3D spaces},''
	\url{https://www.projectaria.com/scenescript/}, accessed: June 2nd, 2024.
	
	\bibitem{fugen2006capability}
	T.~Fugen, J.~Maurer, T.~Kayser, and W.~Wiesbeck, ``Capability of 3-d ray
	tracing for defining parameter sets for the specification of future mobile
	communications systems,'' \emph{IEEE Transactions on antennas and
		propagation}, vol.~54, no.~11, pp. 3125--3137, 2006.
	
	\bibitem{pathak2013uniform}
	P.~H. Pathak, G.~Carluccio, and M.~Albani, ``The uniform geometrical theory of
	diffraction and some of its applications,'' \emph{IEEE Antennas and
		Propagation magazine}, vol.~55, no.~4, pp. 41--69, 2013.
	
	\bibitem{paknys2016uniform}
	R.~Paknys, ``Uniform theory of diffraction,'' 2016.
	
	\bibitem{degli2011analysis}
	V.~Degli-Esposti, V.-M. Kolmonen, E.~M. Vitucci, and P.~Vainikainen, ``Analysis
	and modeling on co-and cross-polarized urban radio propagation for
	dual-polarized mimo wireless systems,'' \emph{IEEE transactions on antennas
		and propagation}, vol.~59, no.~11, pp. 4247--4256, 2011.
	
	\bibitem{tensorflow}
	\BIBentryALTinterwordspacing
	M.~Abadi \emph{et~al.}, ``{TensorFlow}: Large-scale machine learning on
	heterogeneous systems,'' 2015, software available from tensorflow.org.
	[Online]. Available: \url{https://www.tensorflow.org/}
	\BIBentrySTDinterwordspacing
	
	\bibitem{blender}
	\BIBentryALTinterwordspacing
	B.~O. Community, \emph{Blender - a 3D modelling and rendering package}, Blender
	Foundation, Stichting Blender Foundation, Amsterdam, 2018. [Online].
	Available: \url{http://www.blender.org}
	\BIBentrySTDinterwordspacing
	
	\bibitem{hoydis2022sionna}
	J.~Hoydis, S.~Cammerer, F.~A. Aoudia, A.~Vem, N.~Binder, G.~Marcus, and
	A.~Keller, ``Sionna: An open-source library for next-generation physical
	layer research,'' \emph{arXiv preprint arXiv:2203.11854}, 2022.
	
	\bibitem{van2008visualizing}
	L.~Van~der Maaten and G.~Hinton, ``Visualizing data using t-sne.''
	\emph{Journal of machine learning research}, vol.~9, no.~11, 2008.

\end{thebibliography}

\end{document}